\documentclass[useAMS,usenatbib,usegraphicx]{mn2e}

\usepackage{aas_macros}

\defcitealias{Seifried11}{Paper~I}
\defcitealias{Blandford82}{BP82}

\title[Magnetic fields in massive star formation II]{Magnetic fields during the early stages of massive star formation -- II. A generalised outflow criterion}
\author[D. Seifried et al.]
  {D.~Seifried,$^{1,2,3}$\thanks{dseifried@hs.uni-hamburg.de} R.~E.~Pudritz,$^3$ R.~Banerjee,$^1$ D.~Duffin,$^3$ R.S.~Klessen$^2$\\
  $^1$Hamburger Sternwarte, Universit\"at Hamburg, Gojenbergsweg 112, 21029 Hamburg, Germany \\
  $^2$Institut f\"ur Theoretische Astrophysik, Universit\"at Heidelberg, Albert-Ueberle-Str. 2, 69120 Heidelberg, Germany\\
  $^3$Department of Physics $\&$ Astronomy, McMaster University Hamilton, Ontario, Canada}

\date{Accepted 2012 January 20. Received 2011 December 09; in original form 2011 September 16}

\pagerange{\pageref{firstpage}--\pageref{lastpage}} \pubyear{2011}

\begin{document}

\label{firstpage}

\maketitle

\begin{abstract}
Numerical simulations of outflows formed during the collapse of 100 M$_{\sun}$ cloud cores are presented. We derive a generalised criterion from magnetohydrodynamical wind theory to analyse the launching mechanism of these outflows. The criterion is successfully applied to the whole outflow structure and cases with sub-Keplerian disc rotation. It allows us to decide whether an outflow is driven centrifugally or by the toroidal magnetic pressure. We show that quantities such as the magnetic field line inclination or the ratio of the toroidal to poloidal magnetic field alone are insufficient to determine the driving mechanism of outflows. By performing 12 simulations with variable initial rotational and magnetic energies we are able to study the influence of the initial conditions on the properties of outflows and jets around massive protostars in detail. Our simulations reveal a strong effect of the magnetic field strength on the morphology of outflows. In runs with weak fields or high rotational energies, well collimated, fast jets are observed whereas for strong fields poorly collimated, low-velocity outflows are found. We show that the occurrence of a fast jet is coupled to the existence of a Keplerian protostellar disc. Despite the very different morphologies all outflows are launched from the discs by centrifugal acceleration with the toroidal magnetic field increasingly contributing to the gas acceleration further away from the discs. The poor collimation of the outflows in runs with strong magnetic fields is a consequence of the weak hoop stresses. This in turn is caused by the slow build-up of a toroidal magnetic field due to strongly sub-Keplerian disc rotation. The mass and momentum outflow rates are of the order of 10$^{-4}$ M$_{\sun}$ yr$^{-1}$ and 10$^{-4}$ M$_{\sun}$ km s$^{-1}$ yr$^{-1}$, respectively. The mass ejection/accretion ratios scatter around a mean of 0.3 in accordance with observational and analytical results. Based on our results we suggest an evolutionary scenario for the earliest stage of massive star formation in which initially poorly collimated outflows develop which successively get better collimated during their evolution due to the generation of fast jets.
\end{abstract}

\begin{keywords}
 hydrodynamics -- magnetic fields -- methods: numerical -- stars: formation -- stars: massive -- stars: winds, outflows.
\end{keywords}

\section{Introduction}

The formation of massive stars and the evolution of associated protostellar outflows is still a highly debated question~\citep[e.g.][]{Beuther05,Zinnecker07}. It is believed that massive stars form in high-mass molecular cloud cores with masses ranging from about 100 M$_{\sun}$ up to a few 1000 M$_{\sun}$. Such cores have characteristic sizes of a few 0.1 pc and peak densities of up to 10$^6$ cm$^{-3}$~\citep[e.g.][]{Beuther07}. A crucial ingredient for the formation of outflows is the magnetic field in the cloud cores. Its importance can be estimated by the mass-to-flux ratio $\mu$ normalised to the critical mass-to-flux ratio~\citep{Mouschovias76} which is essentially the ratio of gravitational to magnetic energy in the region (see equation~(\ref{eq:mu})). For high-mass star forming cores the observed mass-to-flux ratios are typically slightly supercritical with $\mu \la 5$ \citep{Falgarone08,Girart09,Beuther10,Crutcher10} indicating a significant influence of magnetic fields on the star formation process. In turbulence simulations, however, also higher values of $\mu$ up to 20, i.e. weaker magnetic fields, have been found~\citep[e.g.][]{Padoan01,Tilley07}. In combination with the observed overall slow rotation of cores~\citep{Goodman93,Pirogov03} all necessary ingredients for the formation of protostellar outflows are present. Indeed, there are a growing number of observations of outflows around massive protostellar objects~\citep[see][for recent compilations]{Beuther02b,Zhang05}. The generation of such massive outflows, in particular their underlying driving mechanism and their properties will be the focus of our interest here.

For low-mass star formation ($M_{\rmn{core}}$ $\sim$ 1 M$_{\sun}$), the generation, evolution and properties of protostellar outflows have been studied in great detail over the last years~\citep[e.g.][]{Allen03,Banerjee06,Mellon08,Machida08,Hennebelle08,Hennebelle09,Commercon10,Tomida10,Duffin09,Duffin11}. Despite the intensive research in this field there is still no consensus about what drives the outflows beside the fact that all outflows are magnetically driven. Different results can be found in literature about whether the outflow is driven by centrifugal acceleration~\citep{Blandford82,Pudritz86,Pelletier92} or by the pressure of the toroidal magnetic field~\citep{Lynden-Bell96,Lynden-Bell03}. A possible reason for this might be the different methods used to analyse the outflows. In this work we seek to clarify this problem by deriving a fully self-consistent, generalised criterion from magnetohydrodynamical (MHD) wind theory to determine the underlying driving mechanism. The criterion is applicable in the same way to low- and high-mass protostellar objects.

Numerical studies on the influence of magnetic fields in massive star forming regions have received attention only recently. \citet{Banerjee07} study the formation of a protostar and the associated outflow in its earliest evolution with extremely high spatial resolution observing an outflow consisting of two components. The interplay of magnetically driven outflows and ionising radiation is analysed by \citet{Peters11} who found a sphere-like outflow morphology. In this context we also mention the work of~\citet{Vaidya11} who study the effect of outflow decollimation due to radiation forces, but at much later times than presented here. Examining the influence of initial turbulence, an aspect not considered in this work, \citet{Hennebelle11} find that strong magnetic fields reduce the number of fragments formed during the collapse. In an accompanying paper~\citep[][in the following Paper~I]{Seifried11} we analyse the influence of the initial condition in massive star forming regions on accretion rates and the build-up of protostellar discs. One of the main results of this study are the remarkable similar accretion rates despite large variations in the initial conditions which we attribute to effects of the magnetic field. A second crucial result which will be of particular importance for this paper is the fact that for simulations with strong magnetic fields ($\mu < 10$) only sub-Keplerian protostellar discs can form due to the very efficient magnetic braking. 

In the present study, we systematically analyse the influence of the rotational and magnetic energy on protostellar outflows. We focus on the underlying launching mechanism and how the initial conditions affect global outflow properties like mass, momentum and morphology. In particular, we are able to examine how the development of a two-component outflow~\citep{Banerjee07} or a sphere-like outflow~\citep{Peters11} depends on the initial configuration. For this purpose we use the same simulations as in \citetalias{Seifried11}. We note that the initial conditions are selected in a way to cover a large parameter space in accordance with observations and numerical simulations. With the generalised outflow criterion derived in this work we will show that magneto-centrifugally driven outflows consist of two different regimes described by the MHD wind theory. In the first regime close to the disc and the rotation axis, acceleration is dominated by the centrifugal force, i.e. gas gets flung outwards along the poloidal magnetic field lines, whereas in the second regime further away from the disc $B_\phi$ starts to dominate the acceleration. Furthermore we will also show that despite large morphological differences all outflows observed in this work are launched by centrifugal acceleration and that the differences in collimation are mainly caused by varying hoop stresses.

The paper is organised as follows. In Section~\ref{sec:techniques} the numerical techniques and the initial conditions of the runs performed are described briefly. Next we derive the generalised criterion to determine the outflow driving mechanism (Section~\ref{sec:crit}). The results of our simulations are presented in Section~\ref{sec:results}. First, two representative simulations are discussed in detail also testing the newly developed outflow criterion. Next, we analyse the long term evolution and stability of the outflows before the effect of varying initial conditions on the outflow properties are examined. In Section~\ref{sec:dis} the results are discussed in a broader context and are compared to theoretical, numerical and observational studies before we conclude in Section~\ref{sec:conclusion}.

\section{Numerical techniques and initial conditions} \label{sec:techniques}

In the following we briefly describe the initial conditions and the numerical methods used in our simulations. For a detailed description we refer the reader to~\citetalias{Seifried11}. The 3-dimensional, MHD simulations of collapsing cloud cores are performed with the AMR code FLASH~\citep{Fryxell00} using a MHD solver developed and tested by~\citet{Waagan11}. The maximum spatial resolution in the simulations is $\Delta x$ = 4.7 AU and the Jeans length is resolved everywhere with at least eight grid cells. We use a threshold density of
\begin{equation}
 \rho_{\rmn{crit}} = 1.78 \cdot 10^{-12} \rmn{g\,cm^{-3}}
 \label{eq:rhocrit}
\end{equation}
for sink particle creation~\citep[for further details on the creation criteria see][]{Federrath10}. 

With sink masses of up to $M_{\rmn{max}} \sim$ 4 M$_{\sun}$, the maximal rotation velocities in the discs and outflow velocities directly coupled to them~\citep{Michel69,Pelletier92}
\begin{equation}
 v_{\rmn{out}} \sim v_{\rmn{rot,max}} \sim v_{\rmn{kep,max}} =  \sqrt{\frac{G M_{\rmn{max}}}{\Delta x}}
\label{eq:vrotmax}
\end{equation}
are limited to about 10 km s$^{-1}$.

The thermodynamical evolution of the gas is modelled by a cooling routine of~\citet{Banerjee06b} taking into account dust cooling, molecular line cooling and the effects of optically thick gas. After the launching of the outflows, we introduce a density threshold of $1 \cdot 10^{-15}$ g cm$^{-3}$ within 67 AU around the simulation centre. This helps to limit the Alfv\'enic velocity
\begin{equation}
 v_{\rmn{A}} = \frac{B}{\sqrt{4 \pi \rho}}
\label{eq:alf}
\end{equation}
thereby preventing prohibitively small timesteps. As shown in~\citetalias{Seifried11}, the dynamical influence of this artificial density threshold is negligible.

The simulated cloud cores have a mass of 100 M$_{\sun}$ and a uniform temperature of 20 K. This mass corresponds to about 56 Jeans masses making the whole configuration highly gravitationally unstable. The core has a diameter of 0.25 pc and is embedded in an ambient medium having a 100 times lower density than at the edge of the core and a 100 times higher temperature of 2000 K to assure pressure equilibrium. The core has a density profile\footnote{To avoid unphysically high densities in the interior of the core, we cut off the $r^{-1.5}$-profile at a radius of 0.0125 pc. Within this radius the density distribution follows a parabola $\rho(r) \propto (1-(r/r_0)^2)$.}
\begin{equation}
 \rho(r) \propto r^{-1.5}.
\end{equation}
We note that for the purely hydrodynamical case, this slope defines the transition between systems with low degree of fragmentation (for steeper density profiles) and with high susceptibility for fragmentation (for shallower density profiles) as shown by~\citet{Girichidis11}.

\begin{figure}
 \includegraphics[width=84mm]{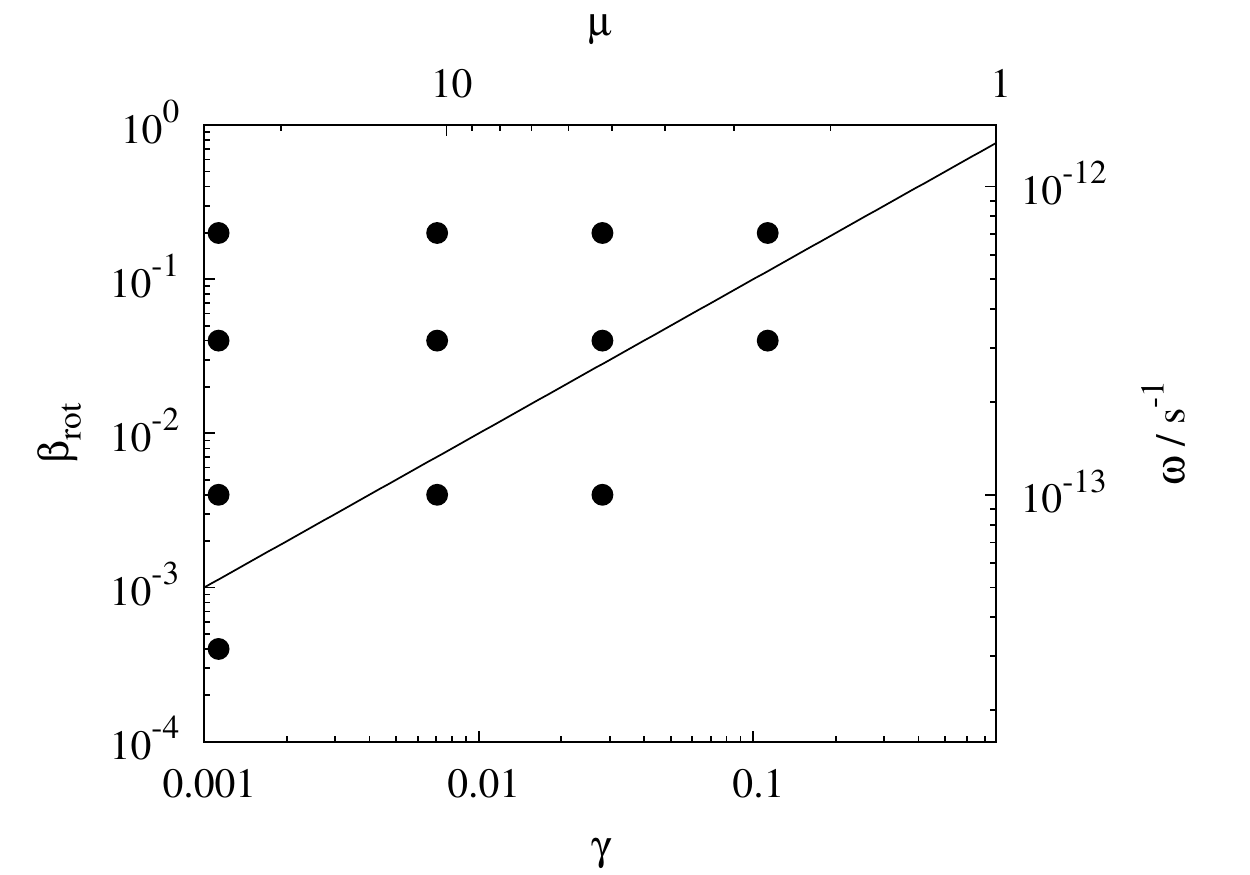}
 \caption{Initial values of the 12 simulations performed. The initial rotational ($\beta_{\rmn{rot}}$) and magnetic energy ($\gamma$) are normalised to the gravitational energy. As can be seen, the starting points bracket the curve where rotational and magnetic energy are equal (black line). The second axes show the corresponding normalised mass-to-flux ratio $\mu$ and the rotation frequency $\omega$.}
 \label{fig:models}
\end{figure}

Whereas the mass distribution described above is kept fixed, the rotational and magnetic energies are changed between the individual simulations. The cores are threaded by a magnetic field with variable strength parallel to the $z$-axis and declining outwards with the cylindrical radius $R$ as
\begin{equation}
 B_z \propto R^{-0.75}\,.
\end{equation}
$B_z$ does not change along the $z$-axis to guarantee $\nabla \bmath{B} = 0$. In addition, the cores are rotating rigidly with the rotation axis parallel to the initial magnetic field. We have performed 12 simulations with varying rotational and magnetic energies. The individual runs with the corresponding parameters are listed in Table~\ref{tab:models}.
\begin{table}
 \caption{Performed simulations with normalised initial mass-to-flux-ratio $\mu$, ratio of magnetic to gravitational energy $\gamma$, magnetic field strength in the centre $B$, ratio of rotational to gravitational energy $\beta_{\rmn{rot}}$, and the corresponding angular frequency $\omega$.}
 \label{tab:models}
 \begin{tabular}{@{}lccccc}
  \hline
  Run & $\mu$  & $\gamma$ & $B$ [$\mu$G] & $\beta_{\rmn{rot}}$ & $\omega$ [10$^{-13}s^{-1}$] \\
  \hline
  26-20 & 26 & $1.13 \cdot 10^{-3}$ & 132 & $2 \cdot 10^{-1}$ & 7.07 \\
  26-4  & 26 &  $1.13 \cdot 10^{-3}$ & 132 & $4 \cdot 10^{-2}$ & 3.16 \\
  26-0.4 & 26 &  $1.13 \cdot 10^{-3}$ & 132 & $4 \cdot 10^{-3}$ & 1.00 \\
  26-0.04 & 26 &  $1.13 \cdot 10^{-3}$ & 132 & $4 \cdot 10^{-4}$ & 0.316 \\
  \hline
  10-20 & 10.4 & $7.06 \cdot 10^{-3}$ & 330 & $2 \cdot 10^{-1}$ & 7.07 \\
  10-4 & 10.4 &  $7.06 \cdot 10^{-3}$ & 330 & $4 \cdot 10^{-2}$ & 3.16 \\
  10-0.4 & 10.4 &  $7.06 \cdot 10^{-3}$ & 330 & $4 \cdot 10^{-3}$ & 1.00 \\
  \hline
  5.2-20 & 5.2 & $2.82 \cdot 10^{-2}$ & 659 & $2 \cdot 10^{-1}$ & 7.07 \\
  5.2-4 & 5.2 &  $2.82 \cdot 10^{-2}$ & 659 & $4 \cdot 10^{-2}$ & 3.16 \\
  5.2-0.4 & 5.2 &  $2.82 \cdot 10^{-2}$ & 659 & $4 \cdot 10^{-3}$ & 1.00  \\
  \hline
  2.6-20 & 2.6 & $1.13 \cdot 10^{-1}$ & 1\,318  & $2 \cdot 10^{-1}$ & 7.07 \\
  2.6-4 & 2.6 &  $1.13 \cdot 10^{-1}$ & 1\,318 & $4 \cdot 10^{-2}$ & 3.16 \\
  \hline
 \end{tabular}
\end{table}
Additionally, the initial values of all runs are shown in Fig.~\ref{fig:models}.
Rotational ($\beta_{\rmn{rot}}$) and magnetic energies ($\gamma$) are normalised to the gravitational energy. As can be seen, both $\beta_{\rmn{rot}}$ and $\gamma$ cover a wide range over more than two orders of magnitude around the line of equipartition, i.e. $\beta_{\rmn{rot}} = \gamma$. For comparative purposes the corresponding mass-to-flux ratio~\citep{Mouschovias76}
\begin{equation}
 \mu = \frac{M_{\rmn{core}}}{\Phi_{\rmn{core}}}/\left(\frac{M}{\Phi}\right)_{\rmn{crit}} = \frac{M_{\rmn{core}}}{\int B_z dA}/\frac{0.13}{\sqrt{G}}\,.
 \label{eq:mu}
\end{equation}
and the angular frequency $\omega$ are listed. We note that the inclusion of turbulent motions in the setup will be delayed to a subsequent paper.

\section{A generalised criterion for magneto-centrifugally driven jets} \label{sec:crit}

Several approaches can be found in literature to determine whether an outflow is driven by centrifugal acceleration~\citep{Blandford82,Pudritz86,Pelletier92} or the magnetic pressure gradient~\citep{Lynden-Bell96,Lynden-Bell03}. A frequently chosen quantity is the ratio of the toroidal to poloidal magnetic field, $B_{\phi}/B_{\rmn{pol}}$~\citep[e.g.][]{Hennebelle08}. If this ratio is significantly above 1, the outflow is often believed to be driven by the magnetic pressure. However, the consideration of $B_{\phi}/B_{\rmn{pol}}$ alone can be misleading as in centrifugally driven flows this value can be as high as 10~\citep[see Fig. 4 in][henceforth BP82]{Blandford82}. Close to the disc surface one can check the inclination of the magnetic field lines w.r.t. the vertical axis. The field lines have to be inclined by more than 30$^{\circ}$ for centrifugal acceleration to work~\citepalias{Blandford82}. Although this criterion is an exact solution of the ideal, stationary and axisymmetric MHD equations for an outflow from a Keplerian disc, its applicability is limited to the surface of the disc. A criterion to determine the driving mechanism above the disc was used by \citet{Tomisaka02} comparing of the centrifugal force F$_{\rmn{c}}$ and the magnetic force F$_{\rmn{mag}}$. By projecting both forces on the poloidal magnetic field lines it can be determined which force dominates the acceleration. For the outflow to be driven centrifugally, F$_{\rmn{c}}$ has to be larger than F$_{\rmn{mag}}$. However, for this criterion to be self-consistent the gravitational force and the fact that any toroidal magnetic field would reduce the effect of F$_{\rmn{c}}$ have to be taken into account.

In the following we derive a generalised criterion based on the ideal, axisymmetric, time-independent (stationary) MHD equations~\citep{Chandrasekhar56,Mestel61} to decide whether centrifugal acceleration dominates or not. The criterion is applicable everywhere in the outflow and not only at the disc surface. To begin with we review the basic results of the solution of the stationary, axisymmetric MHD equations. As the solution is discussed broadly in literature~\citep[e.g.][]{Blandford82,Pelletier92,Spruit96} we only state the main results necessary to derive our criterion. One key aspect of the solution is that it invokes four surface functions which are constant on each magnetic surface, i.e. along each magnetic field line. We list them in the following.
\begin{enumerate}
 \renewcommand{\theenumi}{\arabic{enumi}.}
\item Combining the induction equation, the divergence-free equation and the continuity equation shows that the poloidal velocity is always parallel to the poloidal magnetic field
\begin{equation}
 \rho \mathbfit{v}_{\rmn{pol}} = k \mathbfit{B}_{\rmn{pol}}
\end{equation}
where $k$, the so-called \textit{mass loading}, is the first surface function, i.e. it is constant along any given field line.
\item The second surface function which can be derived from the three aforementioned equations is the \textit{angular velocity} of the magnetic surface
\begin{equation}
 \omega = \frac{v_{\phi}}{r} - \frac{k B_{\phi}}{r \rho} \; .
 \label{eq:omega}
\end{equation}
Here $r$ is the cylindrical radius and $v_{\phi}$ the local rotation speed of the gas.
\item Making use of the equation of motion yields the third surface function, the \textit{angular momentum invariant}
\begin{equation}
 l = r v_{\phi} - \frac{r B_{\phi}}{4 \pi k} \; .
 \label{eq:k}
\end{equation}
\item Finally, considering the kinetic energy equation yields the \textit{energy invariant} along a poloidal field line

\begin{eqnarray}
 \epsilon &=& \frac{1}{2} v^2 + \Phi + h - \frac{r \omega B_{\phi}}{4 \pi k} \nonumber \\
          &=& \frac{1}{2} v_{\rmn{pol}}^2 + \frac{1}{2} v_{\phi}^2 + \Phi + h - \frac{v_{\phi}}{v_{\rmn{pol}}}\frac{1}{4 \pi}\frac{B_{\phi} B_{\rmn{pol}}}{\rho} + \frac{1}{4 \pi}\frac{B_{\phi}^2}{\rho}
 \label{eq:eps}
\end{eqnarray}
where $h$ is the enthalpy and $\Phi$ gravitational potential. This equation is also called the Bernoulli equation. 
\end{enumerate}

In the second line of equation~(\ref{eq:eps}) we have used the variables which can be inferred from simulations to describe $\epsilon$. We especially note the last term containing $B_{\phi}^2$ describing the influence of the toroidal magnetic field on the dynamics of outflows. It states that in magnetic wind theory besides the poloidal magnetic field flinging material outwards, the toroidal magnetic field can contribute to the acceleration of gas~\citep[see also][for a detail discussion]{Spruit96}. Magneto-centrifugal acceleration therefore has two regimes; a centrifugally dominated and a $B_\phi$ (magnetic pressure) dominated regime. In the following we will use \textit{centrifugal} acceleration when describing regions in which gas is accelerated mainly by flinging gas outwards along the poloidal field lines. In contrast, \textit{magneto-centrifugal} acceleration also contains the possibility of $B_\phi$ dominated acceleration.

In the following we always assume that the thermal pressure can be neglected, i.e. $h = 0$. Whether gas can get accelerated, i.e. whether $v_{\rmn{pol}}$ increases, depends on the change of the second to last term on the r.h.s. of equation~(\ref{eq:eps}) (second line) along the poloidal field line where $\epsilon$ is constant. The general condition for acceleration is therefore
\begin{equation}
 \partial_{\rmn{pol}} \left ( \frac{1}{2} v_{\phi}^2 + \Phi - \frac{v_{\phi}}{v_{\rmn{pol}}}\frac{1}{4 \pi}\frac{B_{\phi} B_{\rmn{pol}}}{\rho} + \frac{1}{4 \pi}\frac{B_{\phi}^2}{\rho} \right ) < 0
 \label{eq:crit2}
\end{equation}
where $\partial_{\rmn{pol}}$ denotes the derivative along the poloidal magnetic field. This criterion is universal and can be applied to the whole outflow structure. It describes all regions of gas acceleration including those dominated by the effect of $B_{\phi}$. Furthermore it only contains variables which are directly accessible from the simulations.

Now, we come back to our original aim to work out a criterion for identifying the regions dominated by centrifugal acceleration. To do so, we find it useful to go to a frame rotating with the magnetic field, i.e. with the angular frequency $\omega$. For this purpose the Bernoulli equation (\ref{eq:eps}) is modified as follows
\begin{equation}
 \epsilon' = \epsilon - l \omega = \frac{1}{2} v_{\rmn{pol}}^2 + \frac{1}{2} \left( v_{\phi} - r \omega \right)^2 + \Phi_{\rmn{cg}}
 \label{eq:eps2}
\end{equation}
where $\Phi_{\rmn{cg}}$ is the centrifugal-gravitational potential
\begin{equation}
 \Phi_{\rmn{cg}}(r,z,\omega) = -\frac{G M}{\sqrt{r^2 + z^2}} - \frac{1}{2} \omega^2 r^2
 \label{eq:phicg}
\end{equation}
with $M$ being the mass of the central object and $G$ the gravitational constant. Compared to equation~(\ref{eq:eps}) all magnetic field terms have vanished. This is due to the fact that in the rotating frame the gas flow is strictly parallel to the magnetic field and therefore the Lorentz force vanishes.

We now make the additional assumption that the magnetic field is strong enough to retain a purely poloidal structure, i.e. $B_\phi = 0$, and to force the gas to corotate with the field. Hence, we can set $v_{\phi}/r = \omega$ (compare equation~(\ref{eq:omega})) causing the second term on the r.h.s. of equation~(\ref{eq:eps2}) to vanish. This in turn reduces the question of whether gas can be accelerated to the question of how $\Phi_{\rmn{cg}}$ changes along the field line. $\Phi_{\rmn{cg}}$ decreasing along the field line causes $v_{\rmn{pol}}$ to increase, i.e. the gas gets accelerated and vice versa. We emphasise that this works because gas can only move parallel to the poloidal magnetic field in the stationary case (see equation~(\ref{eq:k})). We note that setting $\omega$ in equation~(\ref{eq:phicg}) to the Keplerian value at the footpoint of the magnetic field line in the disc yields the well-known criterion found by \citetalias{Blandford82} stating that the field lines at the disc surface have to be inclined by more than 30$^{\circ}$ w.r.t. the $z$-axis for centrifugal acceleration to work. However, the expression here is more general and sub-Keplerian velocities may also be used.

In a numerical simulations, for an arbitrarily chosen point somewhere above the disc it is often not possible to determine the footpoint of the magnetic field line passing through that point. Therefore, it is not possible to set $\omega = \Omega_0$ where $\Omega_0$ is the angular frequency at footpoint in the disc where the field line is anchored. Under the assumption that gas and magnetic field corotate, we can circumvent this problem. We therefore replace $\omega$ by the local angular frequency of the gas, i.e. $\omega = \Omega = v_{\phi}/r$ which is known from the simulation data. Hence, at any given point (R$_{\ast}$,Z$_{\ast}$) we can calculate the value of the centrifugal-gravitational potential
\begin{equation}
 \Phi_{\rmn{cg},\ast} = \Phi_{\rmn{cg}}(\rmn{R}_{\ast},\rmn{Z}_{\ast}) = -\frac{G M}{\sqrt{\rmn{R}_{\ast}^2 + \rmn{Z}_{\ast}^2}} - \frac{1}{2} v_{\phi,\ast}^2 \; .
\end{equation}
By solving the equation
\begin{equation}
 \Phi_{\rmn{cg}}(r,z,\omega_{\ast}) = -\frac{G M}{\sqrt{r^2 + z^2}} - \frac{1}{2} \omega_{\ast}^2 r^2 = \Phi_{\rmn{cg},\ast} = const.
 \label{eq:Phicgstar}
\end{equation}
under the assumption that the gas would rotate everywhere with $\omega_{\ast} = v_{\phi,\ast}/R_{\ast}$, one obtains the shape of an isocontour
\begin{eqnarray}
 \lefteqn{z(r,\omega_{\ast},\Phi_{\rmn{cg},\ast}) =} \nonumber \\
 \lefteqn{-\frac{\sqrt{-r^6 \omega_{\ast}^4 - 4 r^4 \omega_{\ast}^2 \Phi_{\rmn{cg},\ast} - 4 r^2 \Phi_{\rmn{cg},\ast}^2 + 4 G^2 M^2}} {r^2 \omega_{\ast}^2 + 2 \Phi_{\rmn{cg},\ast}} \; .}
 \label{eq:contour}
\end{eqnarray}
Along this line the value of $\Phi_{\rmn{cg}}$ stays constant, i.e. $\Phi_{\rmn{cg}} = \Phi_{\rmn{cg},\ast}$. Assume the magnetic field line passing through the point (R$_{\ast}$,Z$_{\ast}$) is rigidly corotating with the gas with the angular velocity $\omega_{\ast}$, one can easily decide whether the gas can get accelerated or not by comparing the shape of the isocontour with that of the magnetic field line. Hence, taking the derivative of equation~(\ref{eq:contour}) w.r.t. $r$ and comparing it to the ratio $B_{z}/B_{r}$ at the point (R$_{\ast}$,Z$_{\ast}$) yields the desired criterion. For centrifugal acceleration to work at an arbitrarily chosen point the criterion therefore is
\begin{equation}
  \frac{r}{z} \frac{1}{G M} \left(\frac{v_{\phi}^2}{r^2}(r^2 + z^2)^{3/2} - GM \right) \left/ \left( \frac{B_{z}}{B_{\rmn{r}}} \right) \right. > 1 \; .
 \label{eq:crit}
\end{equation}
For the sake of simplicity we have neglected the $\ast$ symbol in the above equation. Every quantity has to be taken at the point considered, i.e. at ($r,z$).

We emphasise that the criterion given in equation~(\ref{eq:crit}) traces the regions where acceleration is mainly by the centrifugal force, i.e. flinging out gas along the poloidal field lines. As we have set $B_\phi = 0$, there is no resulting Lorentz force along the poloidal field line~\citep[see also Section 4.1 of][]{Spruit96} and the only acceleration is indeed due to the centrifugal force exceeding the gravitational force. The magnetic field only enforces the gas to rotate with the angular frequency $\omega$. Regions of acceleration not traced by this criterion should be fit by the criterion given in equation~(\ref{eq:crit2}) as here the acceleration by $B_{\phi}$ is taken into account. Therefore, by comparing the results of both criteria we are able to distinguish between regions dominated by centrifugal acceleration and those by the toroidal magnetic pressure.

We again emphasise that the formulation given in equation~(\ref{eq:crit2}) states that in magnetic wind theory, the acceleration by $B_{\phi}$ in magneto-centrifugal flows is implicitly contained. This seems often to be overlooked in literature when discussing the underlying outflow mechanism. Equation~\ref{eq:crit2} nicely demonstrates that a classification into centrifugally driven flows on the one hand and magnetic tower flows on the other might be an oversimplification as in reality there can be a coexistence of centrifugal acceleration and acceleration due to the pressure gradient of $B_{\phi}$. We will refer to this several times in the paper.
We also point out that the equations~\ref{eq:eps} and \ref{eq:eps2} are equivalent despite their very different appearance. The equations~\ref{eq:crit2} and \ref{eq:crit}, however, are not equivalent any more due to the additional assumption of corotation, i.e. $B_\phi = 0$, made for the derivation of equation~(\ref{eq:crit}).

\section{Results} \label{sec:results}

In this section, we present the results of our collapse simulations focussing on the evolution and the launching mechanism of outflows. The evolution of protostellar discs and the accretion properties of the protostars were studied in detail in~\citetalias{Seifried11}. We limit our consideration to the phase after the first sink particle has formed to study the time evolution of the outflow and its underlying launching mechanism in detail. In the following, all times refer to the time elapsed since the formation of the first sink particle.

In general, the outflows are launched shortly after the creation of the first sink particle as soon as a protostellar disc builds up. In each simulation the outflow evolves over time and shows no signs of re-collapse until the end of the simulation. However, outflow morphologies and global properties like mass or momentum (see Table~\ref{tab:mout}) differ significantly between the individual runs. Hence, in the following we select two representative simulations with equal initial rotational energies but different magnetic field strengths to study the properties and the launching mechanism of the outflows in detail.
\begin{table*}
 \caption{Listed are the mass, momentum and extension of the outflows, the total sink particle mass at the end of each simulation, the time the runs have being followed after the first sink particle has formed, and the time averaged mass and momentum outflow rates.}
 \label{tab:mout}
 \begin{tabular}{@{}lccccccc}
  \hline
  Run & $M_{\rmn{out}}$  &   $P_{\rmn{out}}$        & L    & $M_{\rmn{sink}}$  & t$_{\rmn{sim}}$ & $\dot{M}_{\rmn{out}}$      & $\dot{P}_{\rmn{out}}$ \\
      &   (M$_{\sun}$)   & (M$_{\sun}$ km s$^{-1}$) & (AU) &   (M$_{\sun}$)    &   (yr)    & (10$^{-4}$ M$_{\sun}$ yr$^{-1}$) & (10$^{-4}$ M$_{\sun}$ km s$^{-1}$ yr$^{-1}$) \\
  \hline
  26-20   &  0.509  &  0.337  &   707  &  1.85  &  5000 & 1.02 & 0.673 \\
  26-4    &  0.853  &  2.37   &  3215  &  2.65  &  5000 & 1.71 & 4.74 \\
  26-0.4  &  0.526  &  2.23   &  3720  &  3.59  &  5000 & 1.05 & 4.46 \\
  26-0.04 &  0.080  &  0.202  &  856   &  4.26  &  5000 & 0.16 & 0.404\\
  \hline
  10-20   &  1.09   &  1.02   &  1240  &  1.28  &  4000 & 2.73 & 2.54 \\
  10-4    &  0.603  &  0.585  &  669   &  2.23  &  4000 & 1.51 & 1.46 \\
  10-0.4  &  0.164  &  0.177  &  445   &  2.98  &  4000 & 0.41 & 0.442\\
  \hline
  5.2-20  &  0.875  &  1.70   &  2110  &  1.78  &  4000 & 2.19 & 4.25 \\
  5.2-4   &  0.656  &  0.715  &  1128  &  2.28  &  4000 & 1.64 & 1.79 \\
  5.2-0.4 &  0.537  &  0.586  &  1100  &  2.55  &  4000 & 1.34 & 1.46 \\
  \hline
  2.6-20  &  0.116  &  0.281  &  1942  &  1.30  &  3000 & 0.39 & 0.938 \\
  2.6-4   &  0.095  &  0.215  &  1455  &  1.48  &  3000 & 0.32 & 0.717 \\
  \hline
 \end{tabular}
\end{table*}

\subsection{Weak field case 26-4} \label{sec:weak}

\subsubsection{General properties} \label{sec:global26}

\begin{figure}
\centering
 \includegraphics[width=72mm]{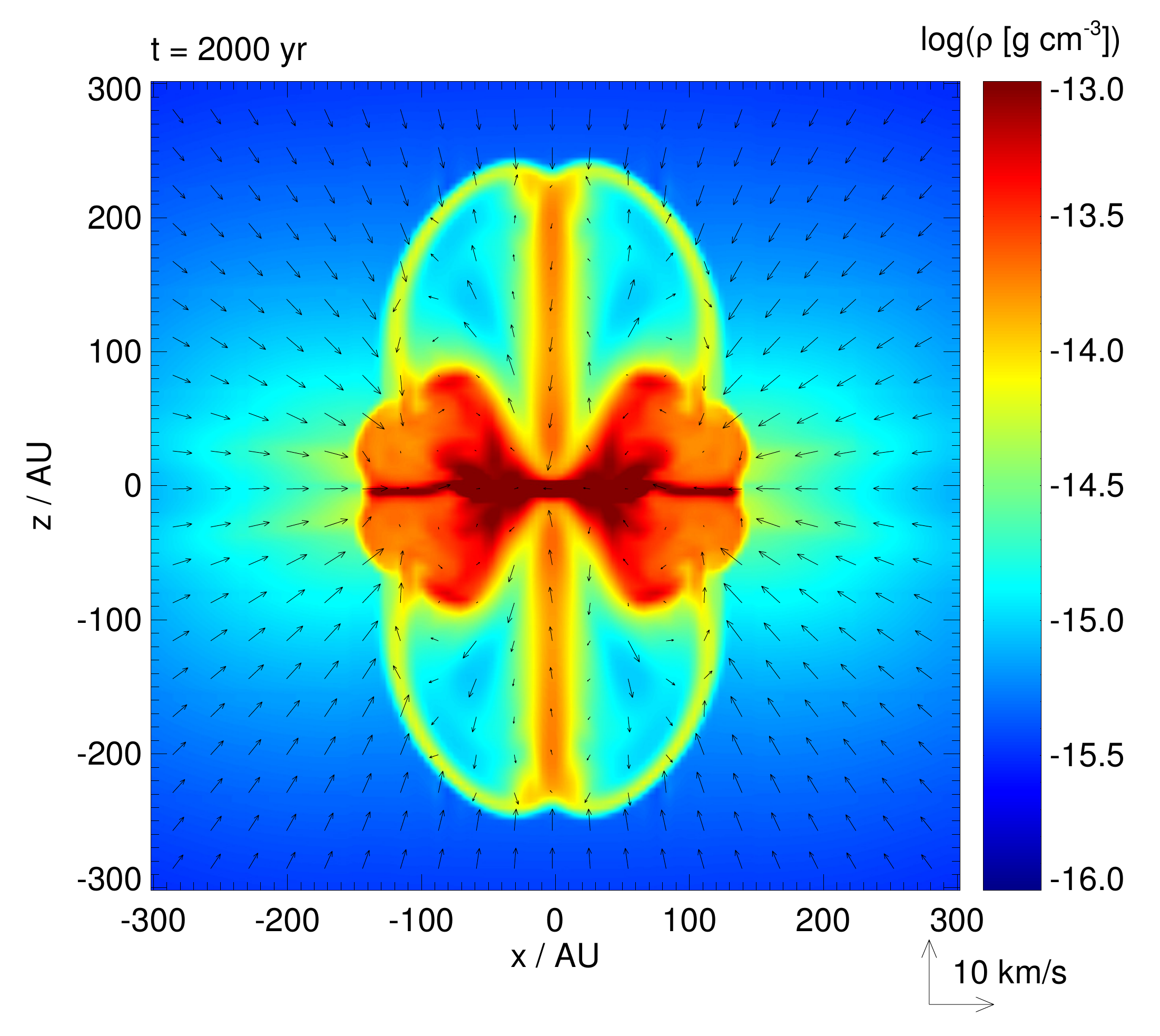}
 \includegraphics[width=72mm]{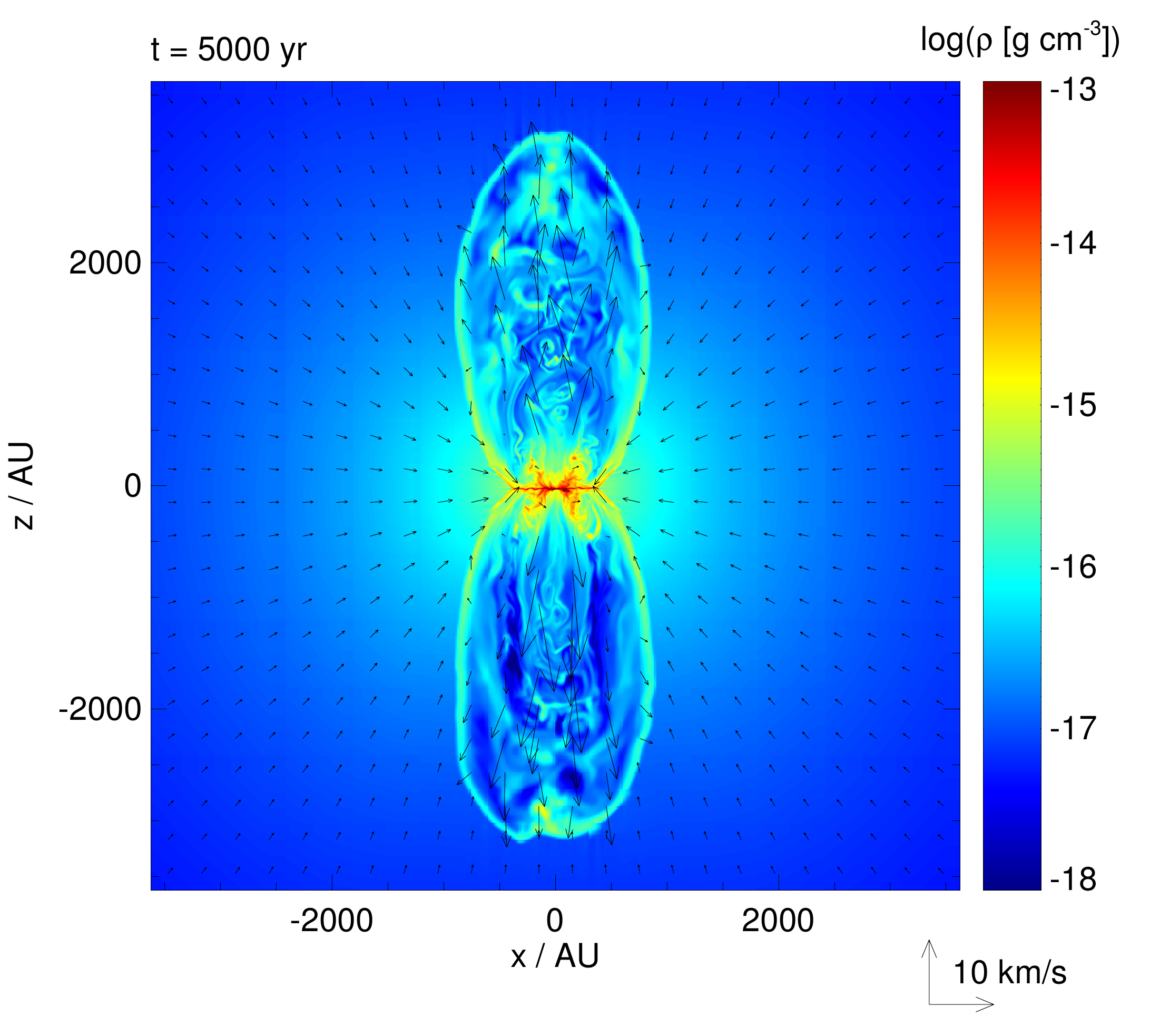}
 \includegraphics[width=72mm]{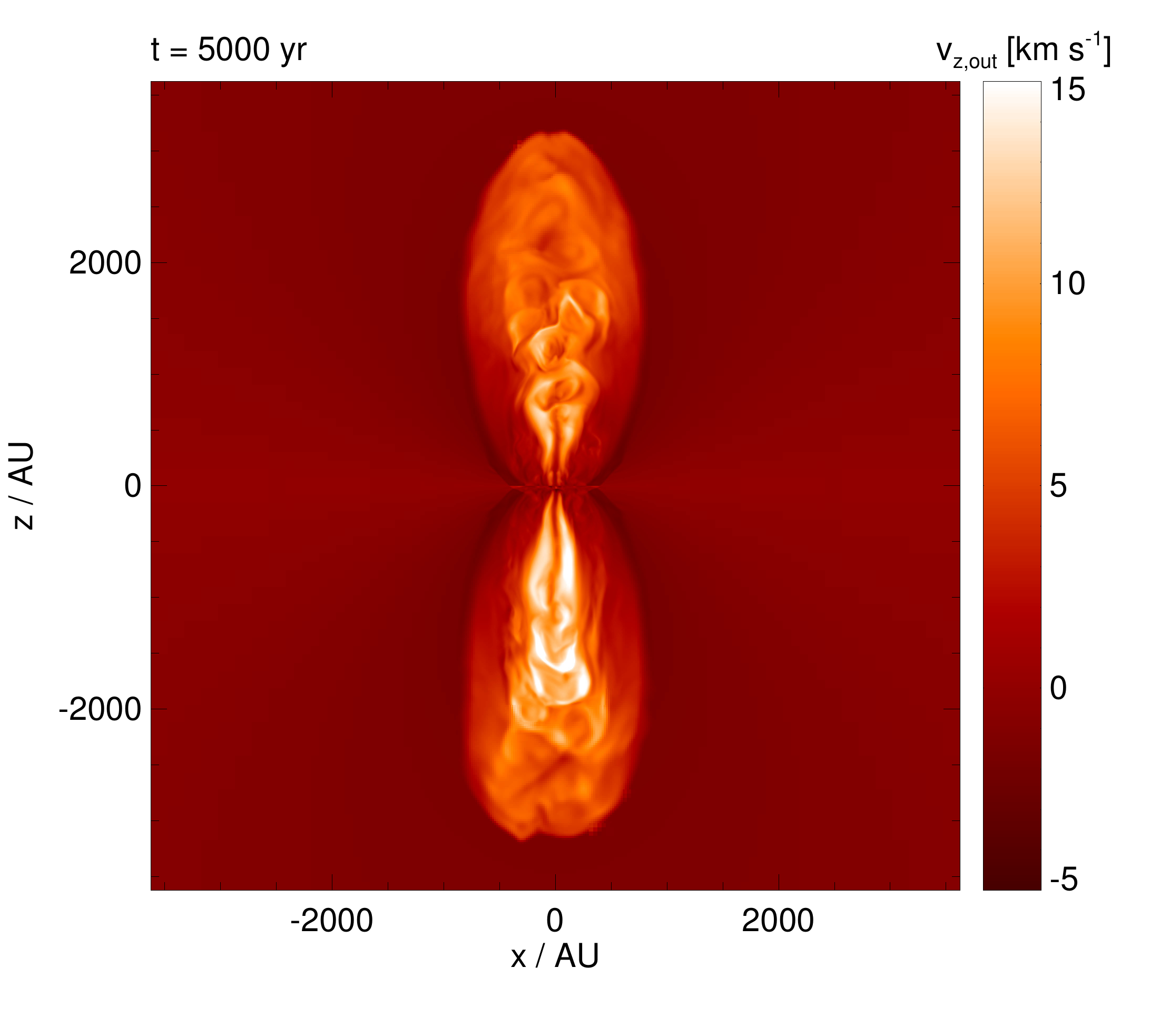}
 \caption{Slice along the $z$-axis in the weak-field run 26-4, at 2000 yr and 5000 yr after the formation of the first sink particle. The two top panels show the density field and the poloidal velocity vectors (black arrows). Note the different spatial scales between the top and middle panel. The outflow velocity ($v_{z,\rmn{out}}$, bottom panel) and the density field after 5000 yr show a very turbulent structure caused by internal shocks.}
 \label{fig:26-4}
\end{figure}

We start our consideration with run 26-4 which has a weak initial magnetic field and a moderate rotational energy (see Table~\ref{tab:models}). In Fig.~\ref{fig:26-4} we show the density, the poloidal velocity field and the outflow velocity
\begin{equation}
 v_{z,\rmn{out}} = v_z \cdot \frac{z}{|z|} 
\end{equation}
 of the outflow in a slice along the $z$-axis at two different times.
It can be seen that initially the outflow expands rather slowly into the surrounding medium as within the first 2000 yr it has reached a height of only $\sim$ 250 AU. After 2000 yr the growth rate increases and reaches a roughly constant value of about 1 AU yr$^{-1} \simeq 4.7$ km s$^{-1}$. We will discuss this behaviour in detail at the end of Section~\ref{sec:launch26}. Fig.~\ref{fig:26-4} also shows that a bow shock at the tip of the outflow develops extending down to the edge of the centrifugally supported disc. As indicated by the velocity vectors, most of the gas within the bow shock is directed vertically outwards with velocities up to 15 km s$^{-1}$. This corresponds to highly supersonic motions of Mach numbers up to $\sim$ 55 using a temperature of 20 K as present in the undisturbed, ambient medium and Mach numbers of about 15 - 20 w.r.t. temperatures of a few 100 K in the outflow itself. The outflow morphology at 2000 yr and 5000 yr reveals a self-similar appearance. In particular the collimation factor of the outflow, i.e. the ratio of the length to the width of one outflow lobe, settles around a value of $\sim$ 4. We especially note the very turbulent structure seen in the outflow after 5000 yr in both the density field and the outflow velocity. Several internal shock fronts and instabilities have occurred in the outflow although no perturbations are included in the initial conditions. This turbulent structure is a consequence of instabilities and not of episodic ejection events. We checked this by visual inspecting the time evolution of the outflow, finding that continuous ejection occurs over the whole time range. Furthermore, a highly time variable ejection rate would also require significant variations in the mass accretion rate~\citep[e.g.][]{Pudritz86} which is clearly not the case~\citepalias[see Fig. 13 in][]{Seifried11}. A turbulent structure as observed here was recently also reported by \citet{Staff10} for jet simulations. The energy in the shocks dissipates and heats up the jet possibly resulting in optical forbidden line emission~\citep{Staff10}. Interestingly, despite the turbulent structure the outflow keeps expanding with constant speed. Another interesting fact is that the gas is continuously ejected from the disc over the whole 5000 yr although the protostellar disc starts to fragment around t $\simeq$ 2500 yr and has formed 12 further fragments by then~\citepalias[see][for details]{Seifried11}. Hence, fragmentation does not necessarily terminate the driving of outflows, a fact which is also observed in run 26-20.

\begin{figure}
 \includegraphics[width=84mm]{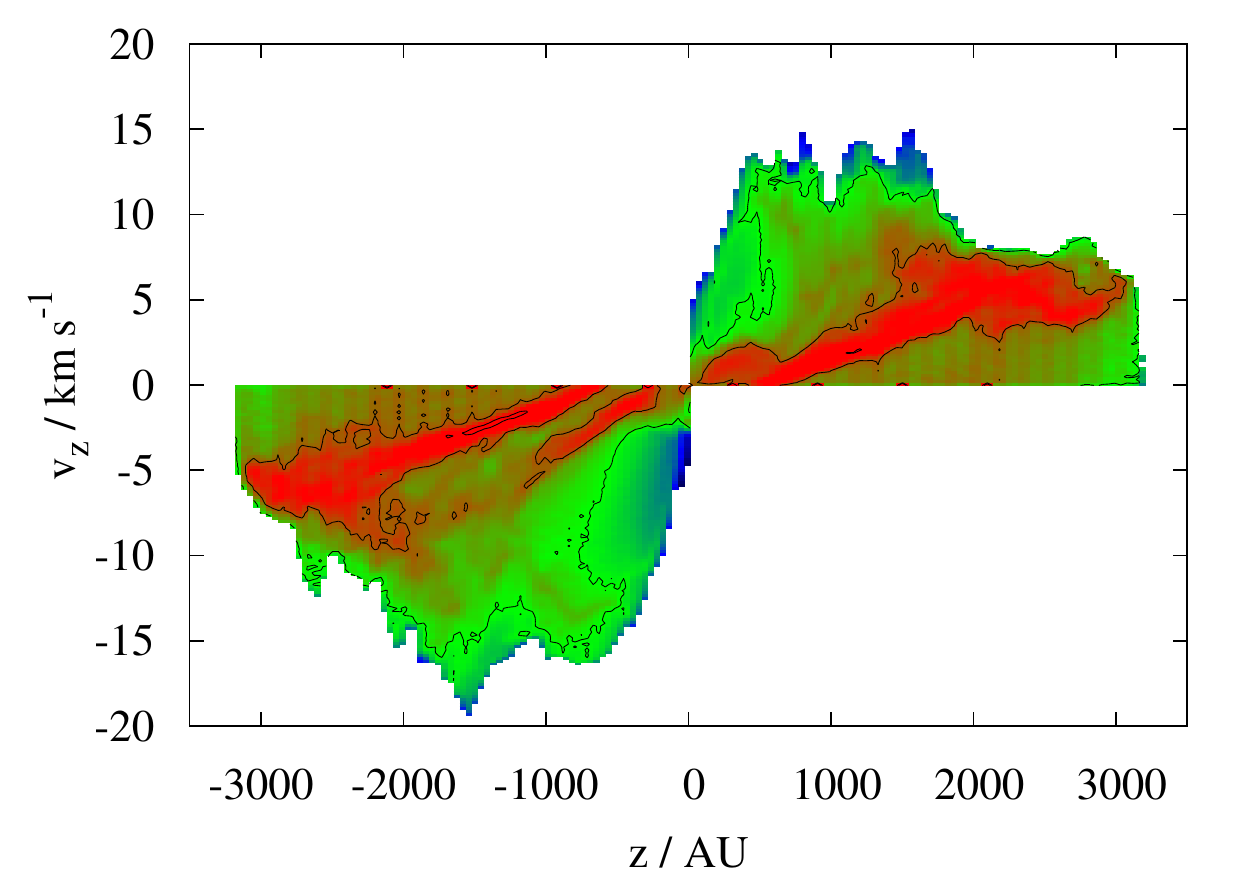}
 \caption{Position-velocity diagram for the weak-field run 26-4 after 5000 yr. The contours have a logarithmic spacing. The bulk velocity increasing with distance. The maximum velocity show several clear peaks which are attributed to internal shock fronts.}
 \label{fig:zpro26-4}
\end{figure}

To study the outflow structure in more detail, we show the position-velocity (PV) diagram of the outflow at 5000 yr in Fig.~\ref{fig:zpro26-4}. The PV diagram is frequently used by observers to study the velocity structure of outflows~\citep[e.g][]{Lada96,Beltran11}.
For the diagram shown here we only take into account outflowing gas. As indicated by the contours, the velocity of the bulk of outflowing material increases with distance from the midplane. Such a ``Hubble Law'' is frequently observed for outflows around low- and also high-mass protostellar objects~\citep[e.g.][]{Lada96,Arce01,Beuther03,Wang11,Ren11,Beltran11}. The maximum outflow velocity saturates around 15 km s$^{-1}$ showing several distinct peaks which we attribute to the internal shocks in the outflow (see bottom panel of Fig.~\ref{fig:26-4}). Above a distance of about 2000 AU, however, the maximum velocity experiences a significant drop from about 15 to 10 km s$^{-1}$. At the same time the ''Hubble law`` describing the evolution of the bulk velocity truncates and the bulk velocity saturates at a value of $\sim$ 5 km s$^{-1}$. We attribute this to a strong internal shock front and the very turbulent structure at these distances which reduce the maximum outflow velocity and prevent an efficient overall acceleration although local gas acceleration is still possible. The position where this happens gradually moves outwards whereas the bulk velocity above this point remains almost constant with a value of $\sim$ 5 km s$^{-1}$.

\begin{figure}
 \includegraphics[width=84mm]{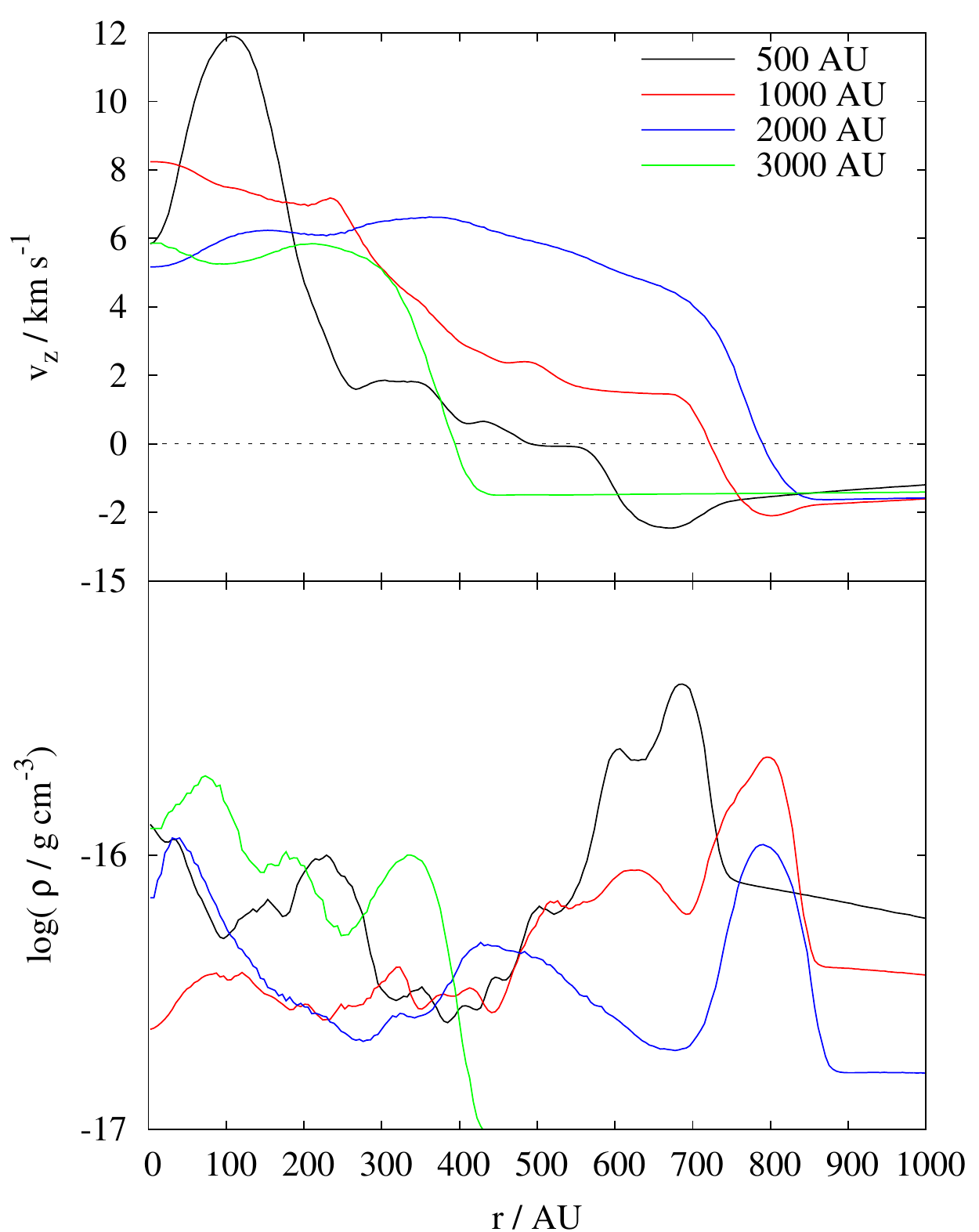}
 \caption{Radial profiles of the outflow velocity (top) and the density (bottom) at different vertical positions for the weak-field run 26-4 after 5000 yr. The quantities are averaged azimuthally before plotting. For $z <$ 2000 AU the outflow velocity increases towards the $z$-axis whereas at larger distances it has an almost flat radial profile. The density profiles are rather flat showing only small variations and a prominent jump associated with the bow shock.}
 \label{fig:rprofile}
\end{figure}

In Fig.~\ref{fig:rprofile} we show radial profiles of the density and outflow velocity at different vertical positions. Both quantities are averaged azimuthally before plotting.
As can be seen, below 2000 AU the highest velocities occur close to the symmetry axis of the jet. In particular the velocity profile at $z$ = 500 AU shows a very well confined, fast velocity component with a quite sharp drop-off at a radius of 150 - 200 AU. This two velocity components, the fast, central as well as the slower, enclosing component are also seen in the PV diagram (Fig.~\ref{fig:zpro26-4}) in particular in the region between $z$ = 0 and $-1000$ AU where two ''Hubble laws'' seem to be present. The good collimation of the fast jet component is most likely due to the strong hoop stress exerted by the toroidal magnetic field. Such two velocity component outflows with a well collimated, fast component close to the axis of the outflow and a wider spread, slower component are frequently observed around low-mass protostars~\citep[see e.g.][for a review]{Bachiller96} and also around massive protostars~\citep[e.g.][]{Beuther04,Ren11}. Furthermore, they are also observed in jet simulations~\citep[e.g.][]{Staff10}.

We note that the decrease of $v_{z,\rmn{out}}$ close to r = 0 at $z$ = 500 AU (black line in the top panel of Fig.~\ref{fig:rprofile}) is most likely a numerical issue as gas very close to the $z$-axis cannot get accelerated properly due to the limited resolution. Above $z$ = 2000 AU the velocity profile is more or less smooth over the whole radial range. In accordance with the PV diagram (Fig.~\ref{fig:zpro26-4}) for those distances no velocities higher than $\sim$ 6 km s$^{-1}$ occur. The density profiles (bottom panel of Fig.~\ref{fig:rprofile}) show a relatively flat shape at all distances with variations of about half an order of magnitude and possibly a slight increase towards the $z$-axis. At the outer edge the density experiences a strong increase due to the bow shock confining the outflow. By carefully comparing with the top panel of Fig.~\ref{fig:rprofile}, it can be seen that the material in the bow shock partly reveals infall velocities.

\subsubsection{Launching mechanism} \label{sec:launch26}

\begin{figure}
 \includegraphics[width=84mm]{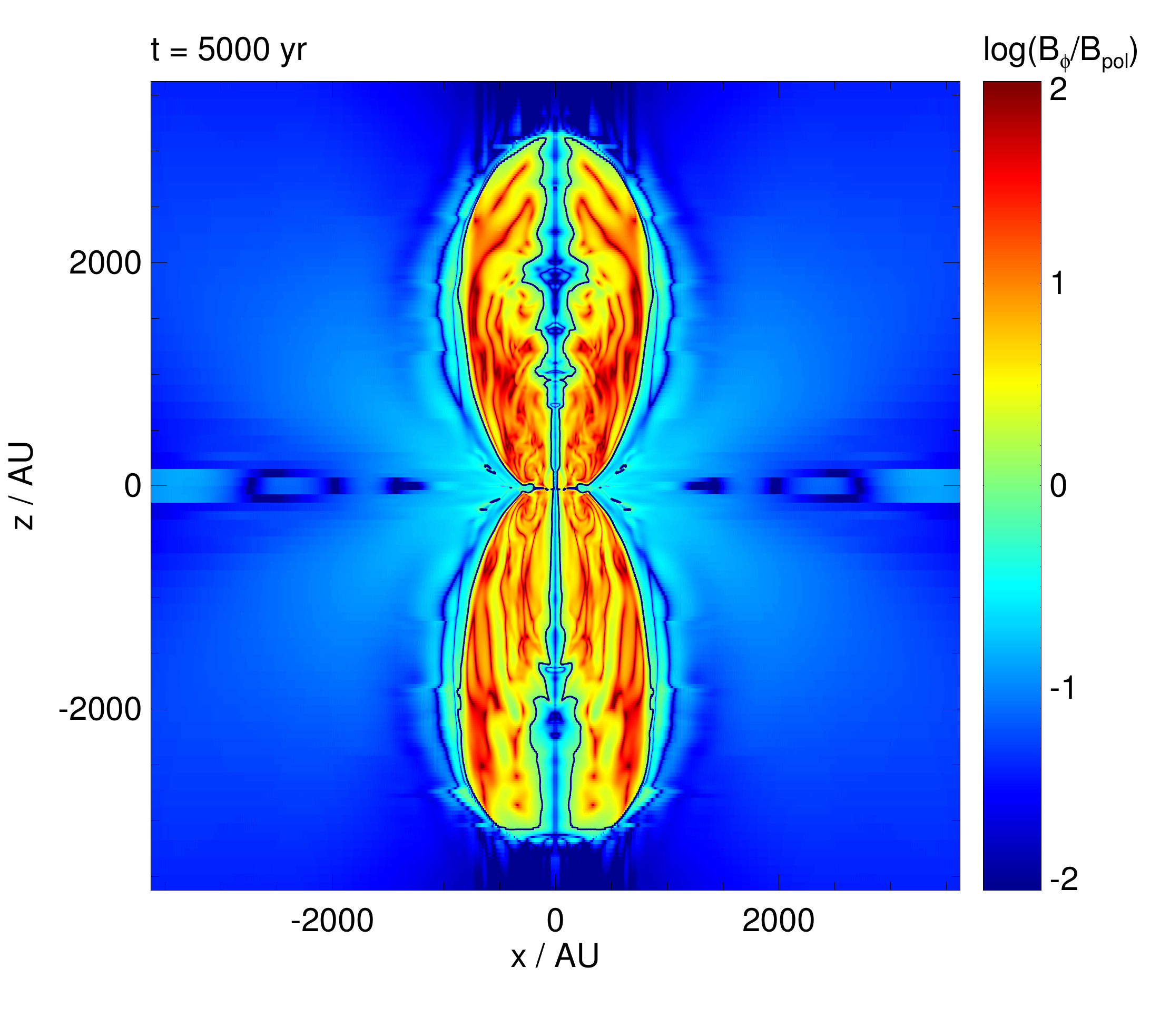} \vspace{-5mm} \\
 \includegraphics[width=84mm]{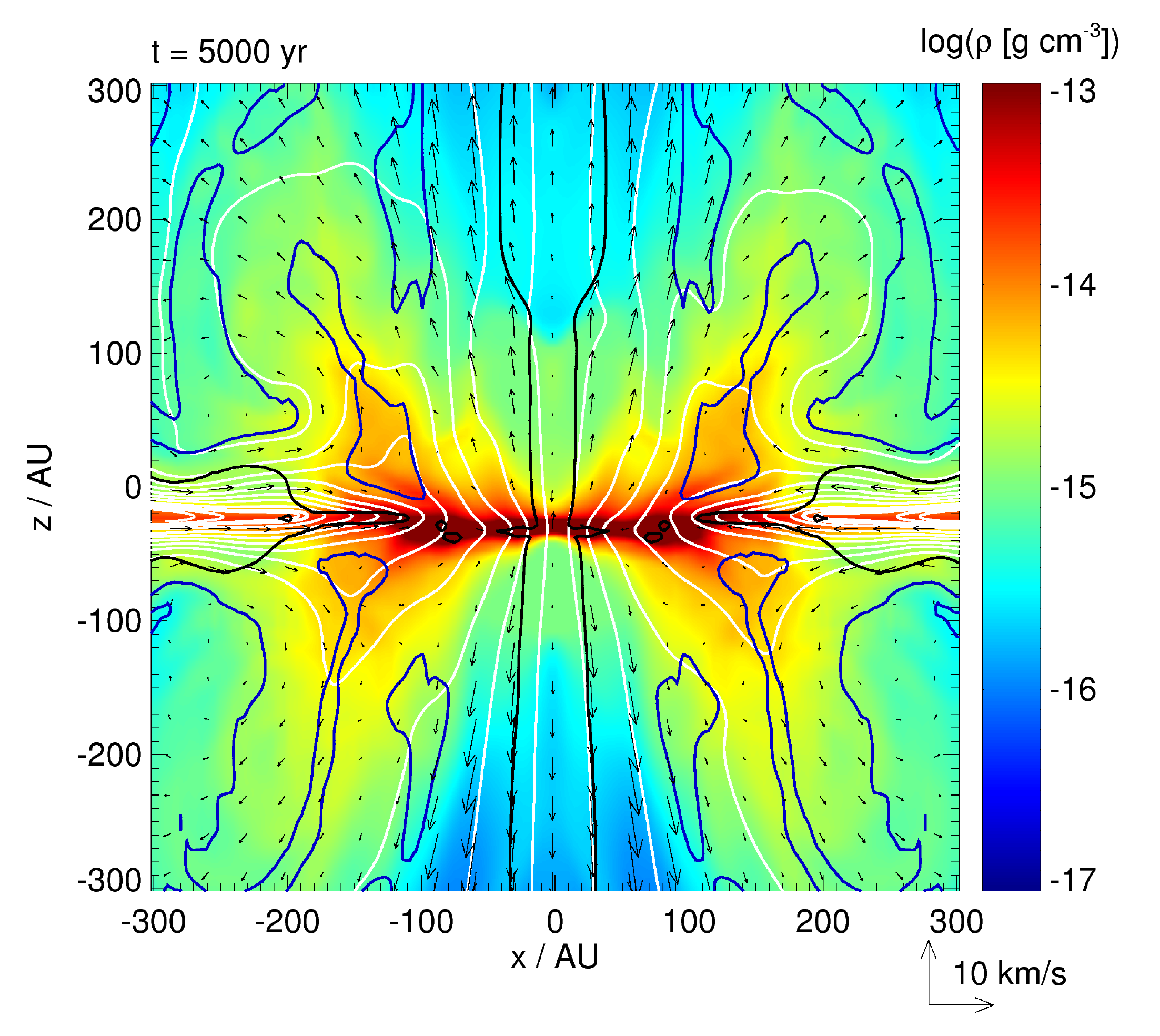}
 \caption{Magnetic field lines structure of the weak-field run 26-4, after 5000 yr in a slice along the $z$-axis. All quantities are averaged azimuthally. Top: Ratio of the toroidal to poloidal magnetic field. Black contours give the transition from the toroidally to the poloidally dominated region. Almost the whole outflow area is dominated by $B_{\phi}$. Bottom: Density field in the inner region. Also shown are the poloidal field lines (white lines), the velocity field (black arrows) and the contours where $B_{\phi}/B_{\rmn{pol}}$ = 1, 10 (black and dark blue lines, respectively). Almost everywhere the poloidal magnetic field has inclination larger than 30$^{\circ}$ w.r.t. the $z$-axis suitable for centrifugal acceleration.}
 \label{fig:mag26-4}
\end{figure}

Next, we analyse the underlying launching mechanism of the outflow in the weak field case 26-4. Once again, in order to smooth out local perturbations which inevitably would occur in an arbitrarily chosen slice along the $z$-axis, the quantities shown in the following are averaged azimuthally. Firstly, the relative importance of the toroidal ($B_{\phi}$) and poloidal magnetic field ($B_{\rmn{pol}}$) for the total magnetic energy content in the outflow is studied. This is done in the top panel of Fig.~\ref{fig:mag26-4} where the value of $B_{\phi}/B_{\rmn{pol}}$ in a slice along the $z$-axis is shown.
As can be seen, almost the complete outflow region is dominated by $B_{\phi}$. The toroidal component is created by the rotation of the protostellar disc in which the initially purely poloidal magnetic field is anchored. $B_{\phi}$ reaches values larger than $B_{\rmn{pol}}$ by up to 2 orders of magnitude. Hence, $B_{\phi}$ should have a crucial effect on the evolution of the outflow. Depending on the position, the absolute values of $B_{\phi}$ and $B_{\rmn{pol}}$ vary between $\sim$ 0.01 G and $\sim$ 1 G.

Outflows driven by the pressure gradient of the toroidal magnetic field are described by \citet{Lynden-Bell96,Lynden-Bell03} and are often termed as magnetic tower flows. The fact that the outflow region is mainly dominated by $B_{\phi}$ suggest that the outflow is such a magnetic tower flow. However, a closer inspection reveals that the region close to the symmetry axis, where the highest velocities occur (see bottom panel of Fig.~\ref{fig:26-4}), is either only weakly dominated by the toroidal magnetic field or even dominated by $B_{\rmn{pol}}$, i.e. $B_{\phi}/B_{\rmn{pol}} \le 10$. We emphasise that even a field configuration which is dominated by $B_{\phi}$ does not contradict the centrifugal acceleration mechanism. \citetalias{Blandford82} find values of $B_{\phi}/B_{\rmn{pol}}$ up to $\sim$ 10 (see their Fig. 4) in the acceleration region in agreement with our findings. Therefore, we assume that the outflow is driven by centrifugal acceleration although there is certainly a contribution to the acceleration by $B_{\phi}$ as well. Of course, this assumption has to be confirmed quantitatively in the following.

To do so, in the bottom panel of Fig.~\ref{fig:mag26-4} we analyse the magnetic field structure in detail concentrating on the inner region where the jet is launched. The poloidal magnetic field lines are overplotted on the density field. In addition, we show the poloidal velocity field and the contours where $B_{\phi}/B_{\rmn{pol}}$ = 1, 10 respectively. As can be seen, the field lines just above/below the disc are strongly inclined against the $z$-axis. Except for the innermost part the inclination angle is everywhere above 30$^\circ$ which is required to launch cold gas from the disc by centrifugal acceleration~\citepalias{Blandford82}. Here we emphasise that an inclination angle smaller than 30$^\circ$ does not necessarily mean that centrifugal launching is impossible. The 30$^\circ$ condition is valid only for a cold gas, i.e. if thermal pressure can be neglected. As soon as a thermal pressure gradient is present aiding to accelerate the gas upwards, even inclination angles below 30$^\circ$ are sufficient for jet launching. As there is clearly a pressure increase towards the centre in the simulation, the innermost region is suitable for centrifugal acceleration as well even though the inclination is below 30$^\circ$. Furthermore, as mentioned before in the inner region where the actual acceleration takes place, the magnetic field is not or only weakly dominated by $B_{\phi}$ in agreement with the findings of~\citetalias{Blandford82}. In contrast, for a magnetic tower flow~\citep{Lynden-Bell03} one would expect a much more wound up structure with a clearly dominating toroidal magnetic field component. However, we note that the toroidal magnetic pressure also contributes to the acceleration of the gas as implied in the MHD wind theory (see equation~(\ref{eq:crit2})). As argued before, the launching from the disc itself, however, is most likely due to centrifugal acceleration.

\begin{figure*}
 \centering
 \includegraphics[width=72mm]{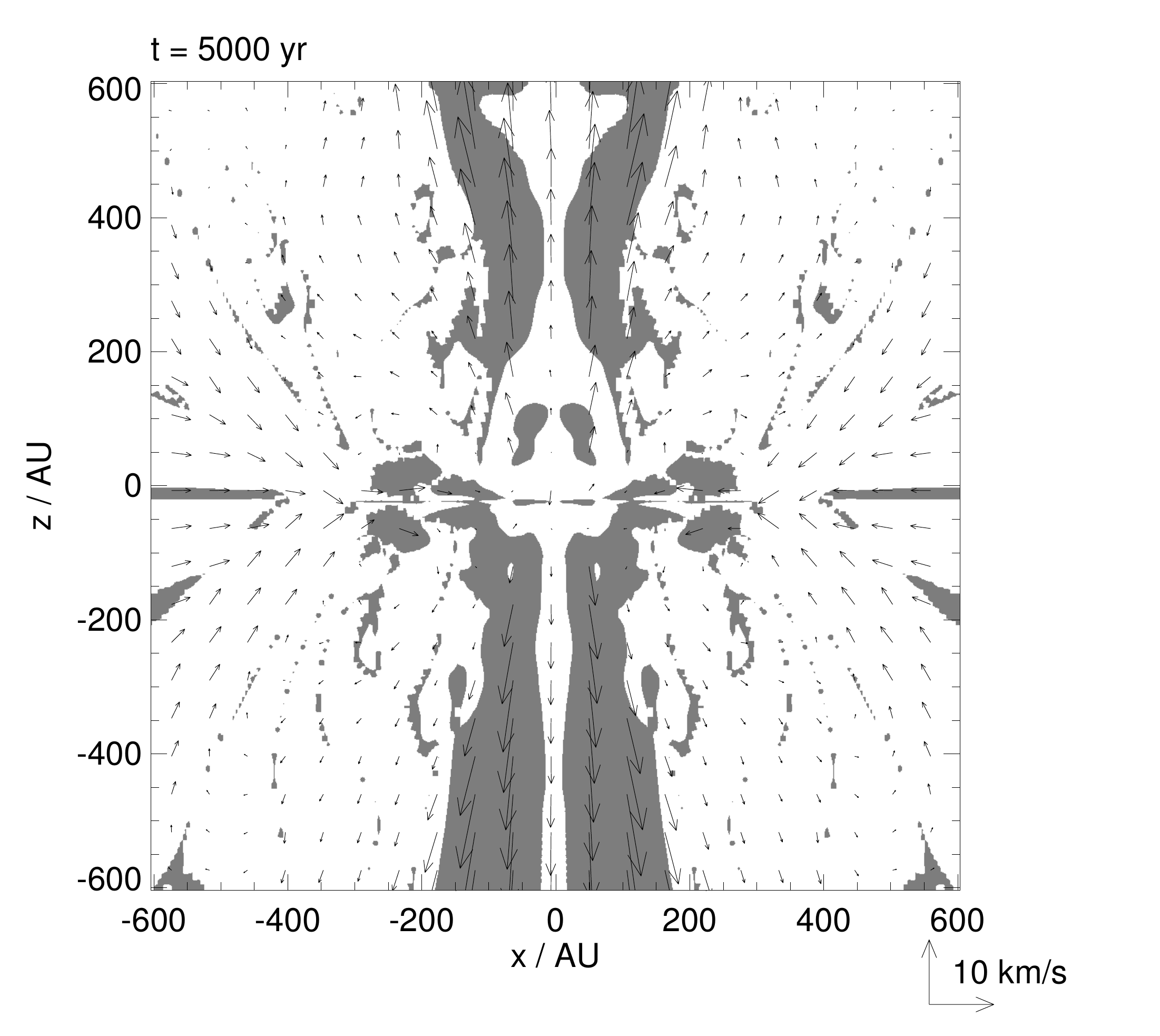}
 \includegraphics[width=72mm]{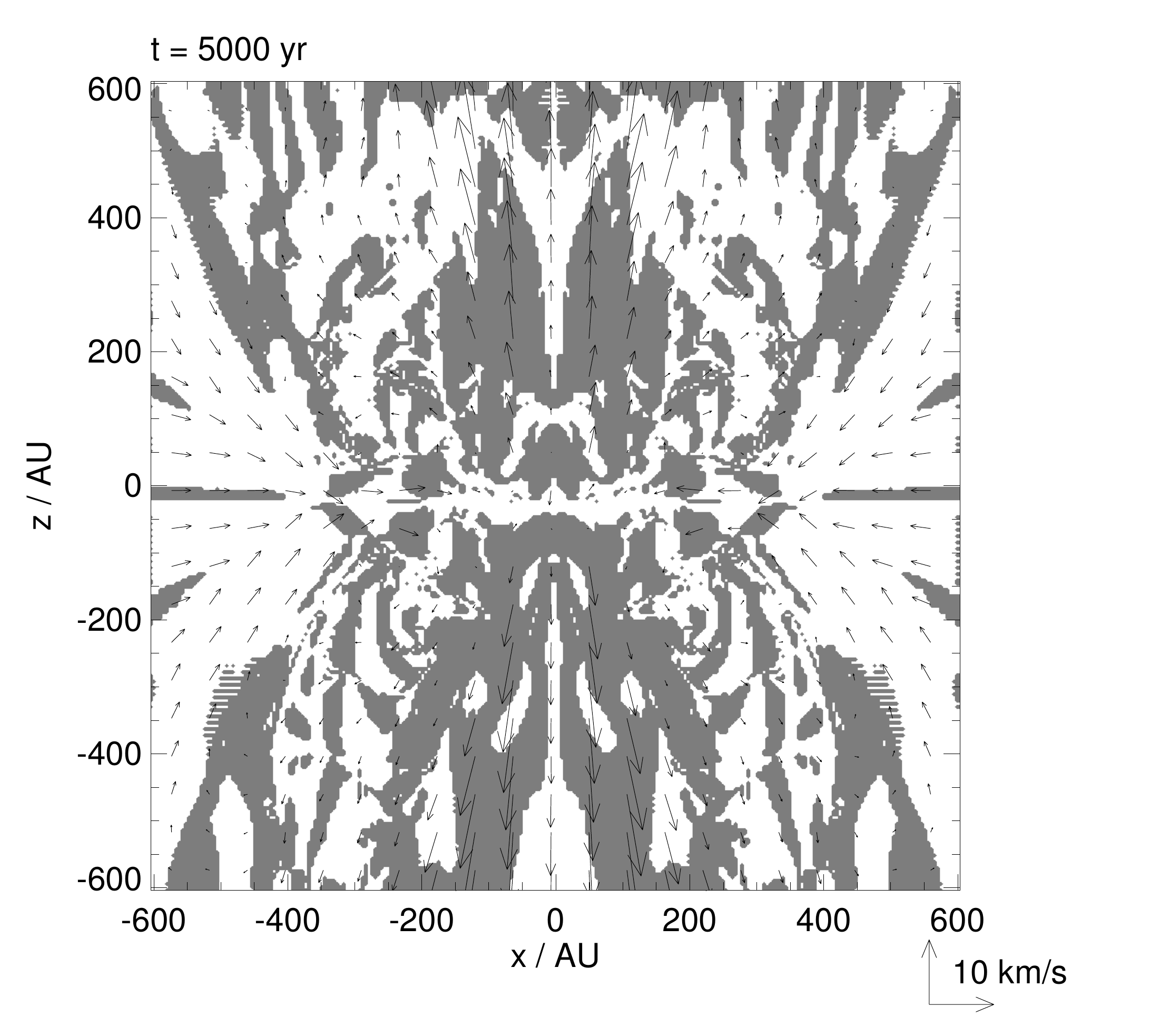} \\
 \includegraphics[width=72mm]{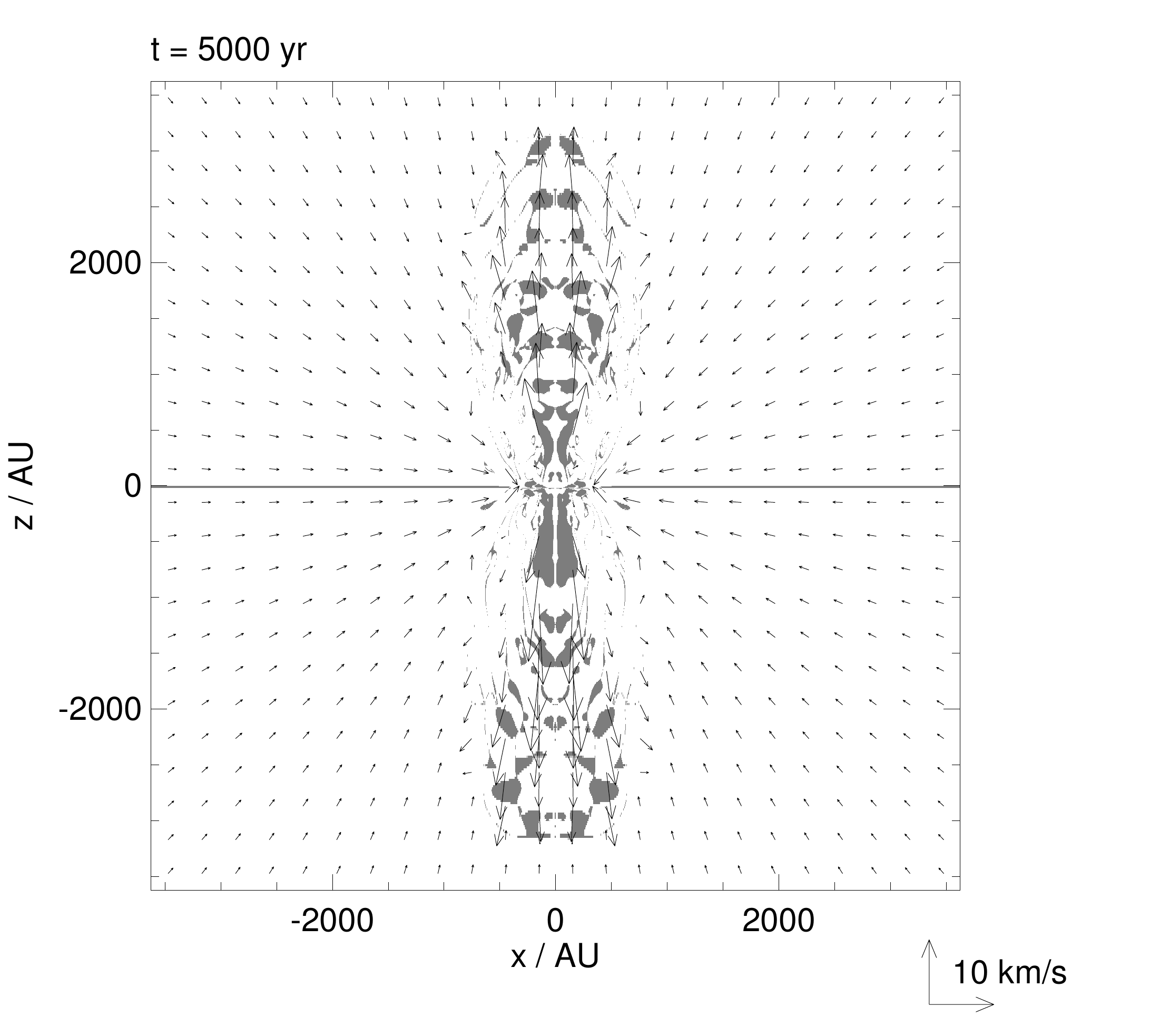}
 \includegraphics[width=72mm]{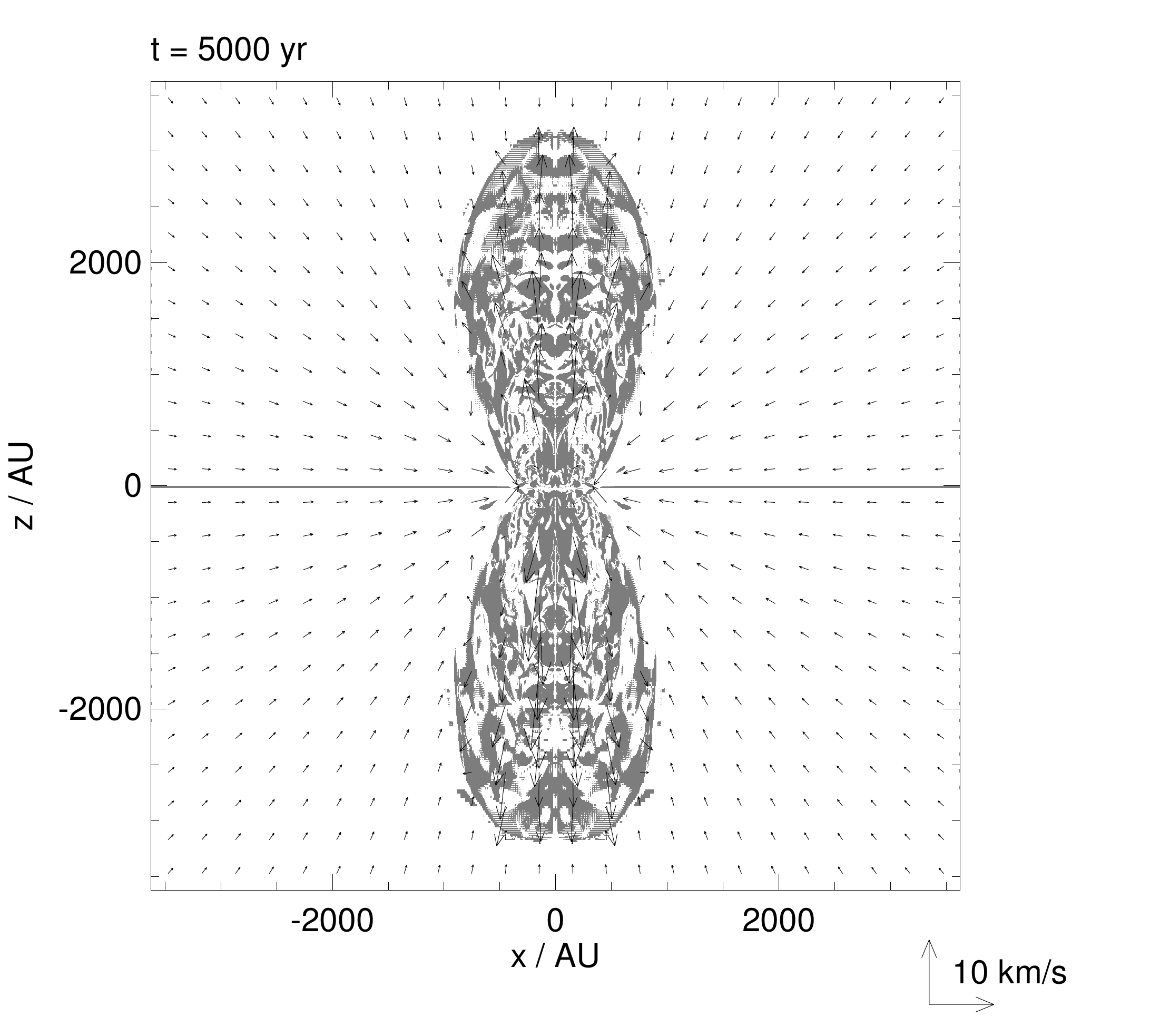}
\caption{Slice along the $z$-axis in the weak-field run 26-4 after 5000 yr for two different scales. Gray shaded areas show the regions where the criteria derived in Section~\ref{sec:crit} (equation~(\ref{eq:crit}) (left panel) and equation~(\ref{eq:crit2}) (right panel)) are fulfilled. The left panels show that centrifugal acceleration works mainly close to the $z$-axis up to a height of about 800 AU which agrees very well with the region where the highest velocities are found. The general criterion is more volume filling and traces also regions in the outer parts.}
 \label{fig:BP82_26}
\end{figure*}

To further support the conclusion of a centrifugally driven wind, we apply the two criteria derived in Section~\ref{sec:crit} (see equations~\ref{eq:crit2} and \ref{eq:crit}). In contrast to the \citeauthor{Blandford82} criterion, we can in general determine the driving mechanism of the outflow away from the disc surface. The results are shown in Fig.~\ref{fig:BP82_26}, where the regions where the criteria are fulfilled are shaded grey.
As can be seen, within about 800 AU above/below the disc there is a continuous region close to the $z$-axis where centrifugally dominated acceleration is possible (left panel). Also at larger heights centrifugal acceleration partly works, in particular close to the symmetry axis of the jet. The grey shaded region around the $z$-axis in the upper left panel of Fig.~\ref{fig:BP82_26} agrees very well with the region where the highest outflow velocities are detected. Hence, despite the strong assumptions like stationarity, axis-symmetry and corotation used to derive the criterion given in equation~(\ref{eq:crit}), it works reasonably well. In particular, it is applicable to regions far from the disc surface in contrast to the 30$^\circ$ - criterion of \citetalias{Blandford82}. We therefore draw the following conclusions: The assumption of corotation is reasonably well fulfilled in the inner part of the outflow ($|z| \le 800$ AU) despite the partly significant toroidal magnetic field (see Fig.~\ref{fig:mag26-4}). Hence, the central jet can be considered as centrifugally driven, i.e. gas gets flung out- and upwards along the poloidal magnetic field lines. The result also demonstrates that the 30$^\circ$ condition -- apart from the disc surface -- and the ratio of $B_{\phi}$ to $B_{\rmn{pol}}$ are not sufficient to determine the driving mechanism above/below the disc.

Taking into account the effect of $B_{\phi}$ in the most general expression of magneto-centrifugal driving (equation~(\ref{eq:crit2})) gives a markedly different result (see right panel of Fig.~\ref{fig:BP82_26}). The grey shaded areas are now much more volume filling, in particular in the outer parts of the outflow with radii $\ga 200$ AU where outflowing gas is present as well. As purely centrifugal dominated acceleration does not work in this region, we expect $B_{\phi}$ to be mainly responsible for the outflow driving. The results in the upper panel of Fig.~\ref{fig:BP82_26} nicely demonstrate the capability of our criterion to distinguish between the different driving mechanisms which was the reason for its derivation. An increasing importance of $B_{\phi}$ for the driving can also be seen in the upper parts of the outflow ($|z| \ge 800$ AU). Here the situation is less suitable for pure centrifugal acceleration (bottom left panel of Fig.~\ref{fig:BP82_26}) but in general magneto-centrifugal acceleration is still possible in great parts (bottom right panel). This is why we argue that $B_{\phi}$ must contribute significantly to the outflow dynamics at great heights.

We note that even the general criterion given in equation~(\ref{eq:crit2}) does not cover the entire outflow structure. This is due to the fact that the gas indeed does not get accelerated everywhere, e.g. in the shock regions, but possibly also due to the azimuthal averaging process. However, the very different results given by the two criteria, in particular in the upper panel of Fig.\ref{fig:BP82_26}, strongly indicate that the two distinct outflow regions are real.

Although the gas in the outflow experiences several internal shocks (see Fig.~\ref{fig:26-4}), the bulk velocity steadily increases within about 2000 AU (Fig.~\ref{fig:zpro26-4}). This can be explained by the fact that acceleration is possible over almost the whole outflow extension (bottom right panel of Fig.~\ref{fig:BP82_26}). After experiencing a shock which decreases the outflow speed, the gas gets reaccelerated again by centrifugal acceleration and/or the toroidal magnetic field pressure. This happens repeatedly over the whole outflow extension thereby successively increasing the bulk velocity. This situation markedly differs from episodic jet ejection which also would produce internal shock fronts. Therefore, we conclude that the knotty structure often observed in protostellar jets is not necessarily a consequence of several outflow ejection events but can also result from a continuously fed jet where gas repeatedly shocks and reaccelerates~\citep[see also][]{Staff10}.

As already seen in the bottom panel of Fig.~\ref{fig:mag26-4} the magnetic field lines get straightened very quickly above and below the disc. We attribute this to the hoop stress produced by the toroidal magnetic field collimating the outflowing gas and therefore also the magnetic field lines. Nevertheless, the gas still gets accelerated centrifugally despite the almost vertical direction of the magnetic field lines (Fig.~\ref{fig:BP82_26}). Furthermore, as shown in Fig.~\ref{fig:mag26-4}, the largest part of the outflow is dominated by the toroidal magnetic field component. Therefore it is not surprising that over its complete extension the outflow as a whole stays well collimated. 

As mentioned in the previous section, the expansion speed is not constant over time but experiences a relatively sharp increase after about 2000 yr. We attribute this to a change in the underlying driving mechanism of the outflow. Indeed, analysing the outflow with our criterion given in equation~(\ref{eq:crit}) shows that within the first 2000 yr purely centrifugal acceleration is not possible indicating that the outflow is mainly driven by the toroidal magnetic pressure. The expansion speed in this phase is almost the same in the vertical and horizontal direction (see top panel of Fig.~\ref{fig:26-4}) with the outer edge of the bubble coinciding with the position of the accretion shock at the disc edge. It is only in this initial stage when we call the outflow a magnetic tower flow~\citep{Lynden-Bell96,Lynden-Bell03}. In contrast to the situation at 5000 yr in this transient phase there is no acceleration of gas from the disc. In fact, the gas is accelerated only at the tip of the outflow. Therefore the situation differs significantly from the later stages. After $\sim$ 2000 yr a fast, well collimated outflow component, the centrifugal driven jet develops in the region close to the $z$-axis. The launching of the jet coincides with the build-up of a well defined, extended ($\sim$ 100 AU) Keplerian disc whereas prior to that disc rotation is mostly sub-Keplerian. 

In summary, besides the magnetic field line structure, the application of the criterion derived in this work strongly indicates that the outflow is mainly driven centrifugally at $|z| \le 800$ AU while the dynamics of $B_{\phi}$ gets more important at large radii and larger heights where the flow is magneto-centrifugally driven.

\subsection{Strong field case 5.2-4} \label{sec:strong}

\subsubsection{General properties}

\begin{figure}
\centering
 \includegraphics[width=72mm]{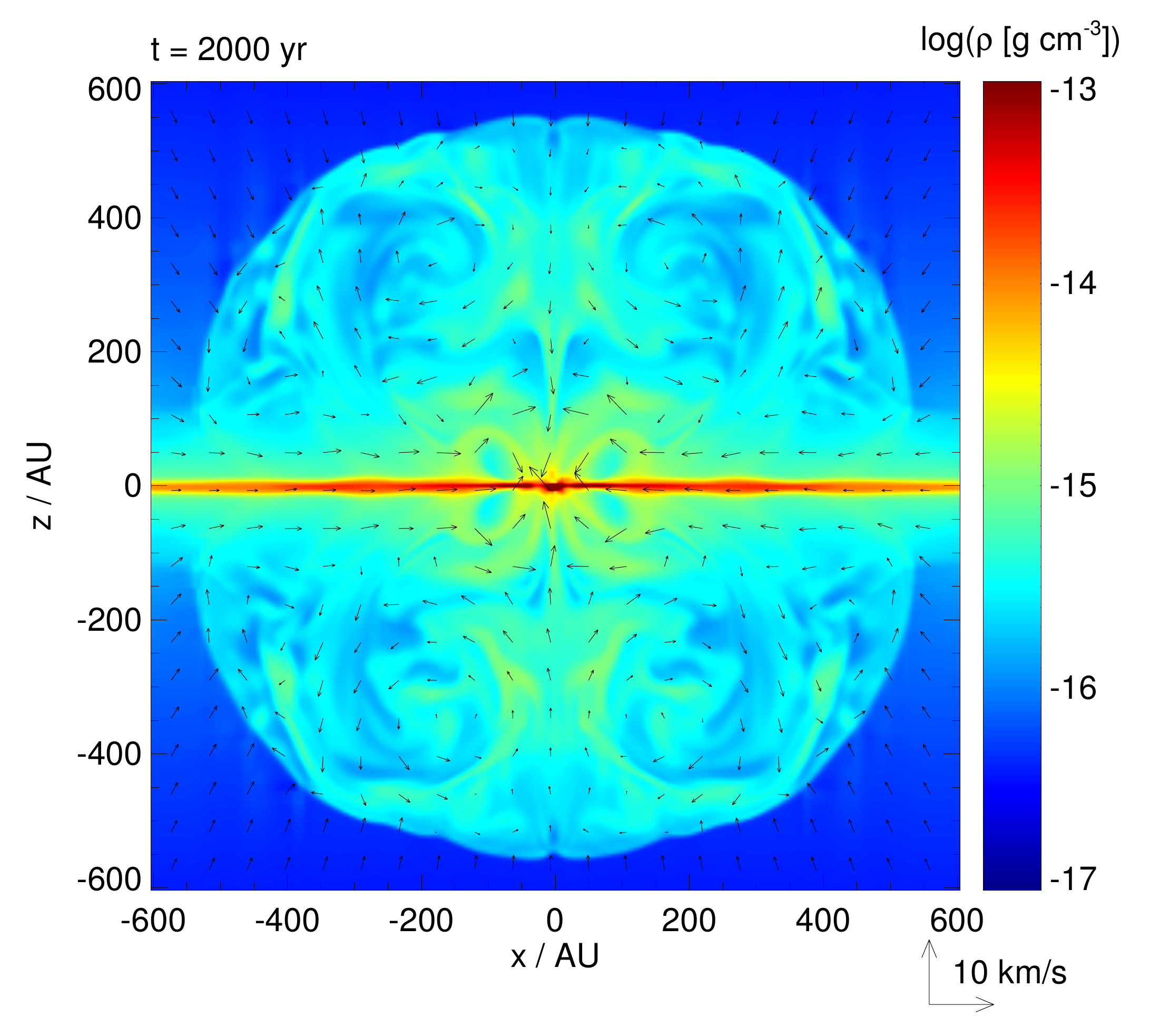}
 \includegraphics[width=72mm]{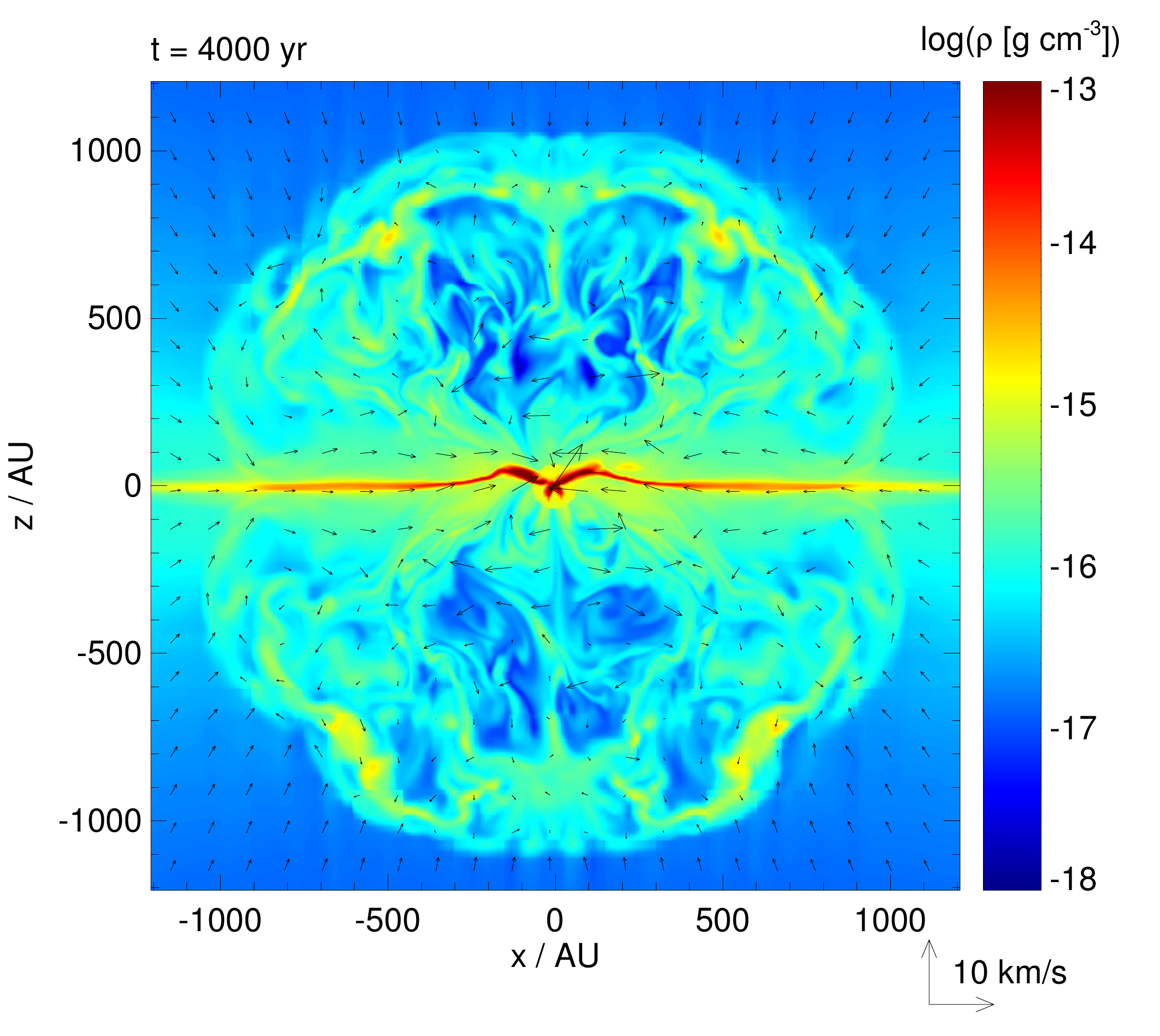}
 \includegraphics[width=72mm]{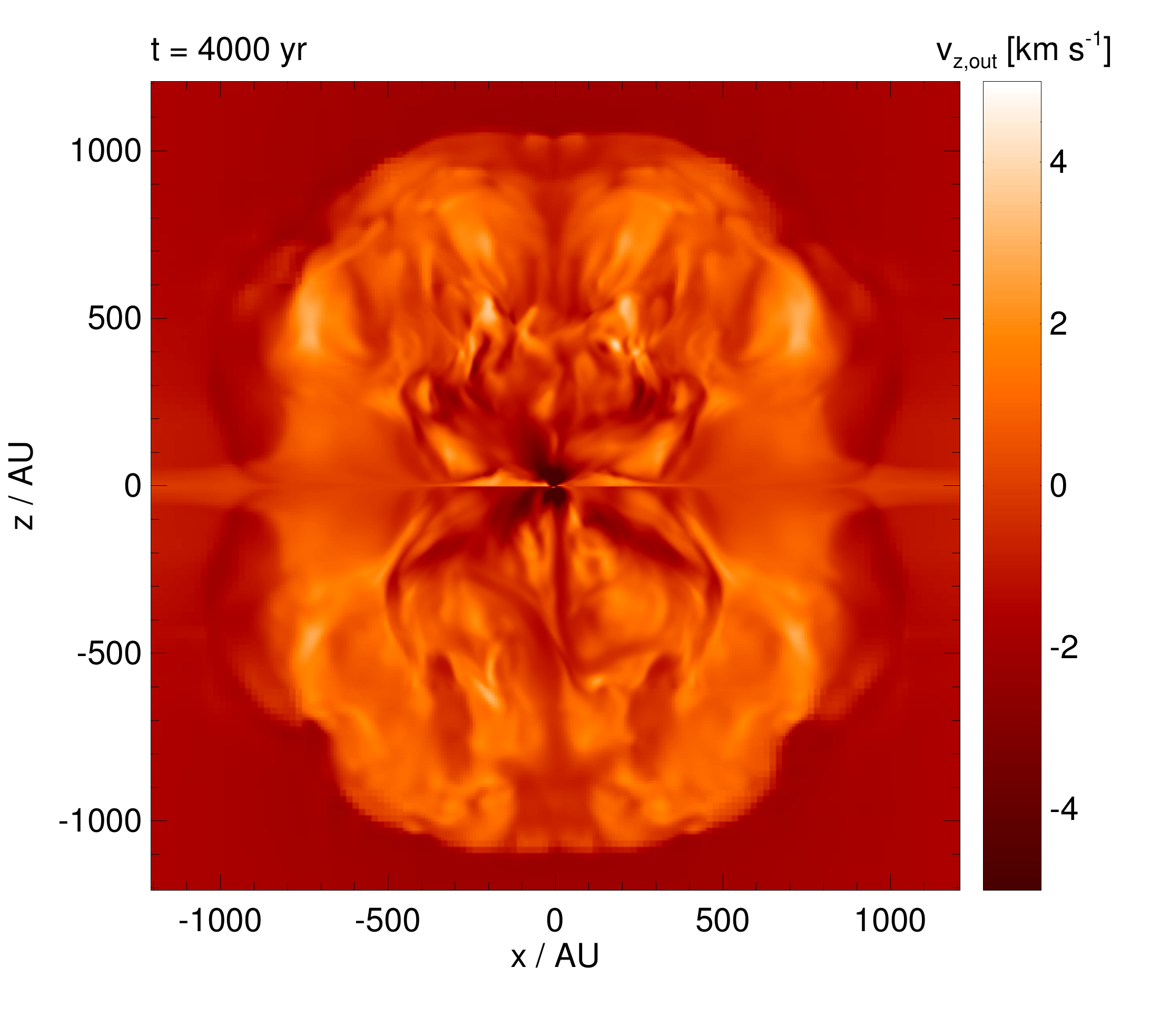}
 \caption{Same as in Fig.~\ref{fig:26-4} but for the strong-field run 5.2-4. The outflow is poorly collimated and has significantly lower outflow velocities than the outflow in run 26-4.}
 \label{fig:5.2-4}
\end{figure}

Next, we describe global properties of the outflow generated in run 5.2-4 which has a 5 times stronger initial magnetic field than run 26-4 (see Table~\ref{tab:models}). The outflow shown in Fig.~\ref{fig:5.2-4} reveals significant differences compared to the outflow in run 26-4 (compare Fig.~\ref{fig:26-4}).
Whereas the latter is well collimated with a collimation factor of $\sim$ 4, the former has a rather sphere-like morphology expanding with roughly the same speed in all directions, therefore also maintaining a self-similar morphology for all times. The outflow velocities reach values of up to 5 km s$^{-1}$ about a factor of 2 - 3 lower than in run 26-4. The expansion speed of the outflow is almost constant over time with a value of 0.28 AU yr$^{-1} \simeq 1.3$ km s$^{-1}$. This is noticeably smaller than the expansion speed of 1 AU yr$^{-1}$ observed in run 26-4. Consequently also the bow shock structure in run 5.2-4 is less pronounced.

Furthermore, a closer inspection of the outflow in run 5.2-4 reveals that in particular close to the symmetry axis and the centre of the bubble gas is still falling inwards even at late times. Gas with outwards motion occurs mainly in the outer wings. The outflow direction in the inner part is almost radial and gets collimated at relatively large radii of $\sim$ 500 AU. This is remarkably different to the situation in run 26-4 where almost all the gas within the outflow area is moving outwards and preferentially parallel to the $z$-axis. A consequence of the complicated velocity structure observed in Fig.~\ref{fig:5.2-4} is the complex density structure showing several shock-like features in the bubble. We find that the flattened structure in the midplane is a strongly sub-Keplerian disc with significant infall motions~\citepalias[see][for a detailed discussion]{Seifried11}. The sub-Keplerian rotation is a consequence of the strong initial magnetic field decelerating the rotation of the initial core via the magnetic braking mechanism~\citep{Mouschovias80}.

\subsubsection{Launching mechanism} \label{sec:launch52}

\begin{figure}
 \includegraphics[width=84mm]{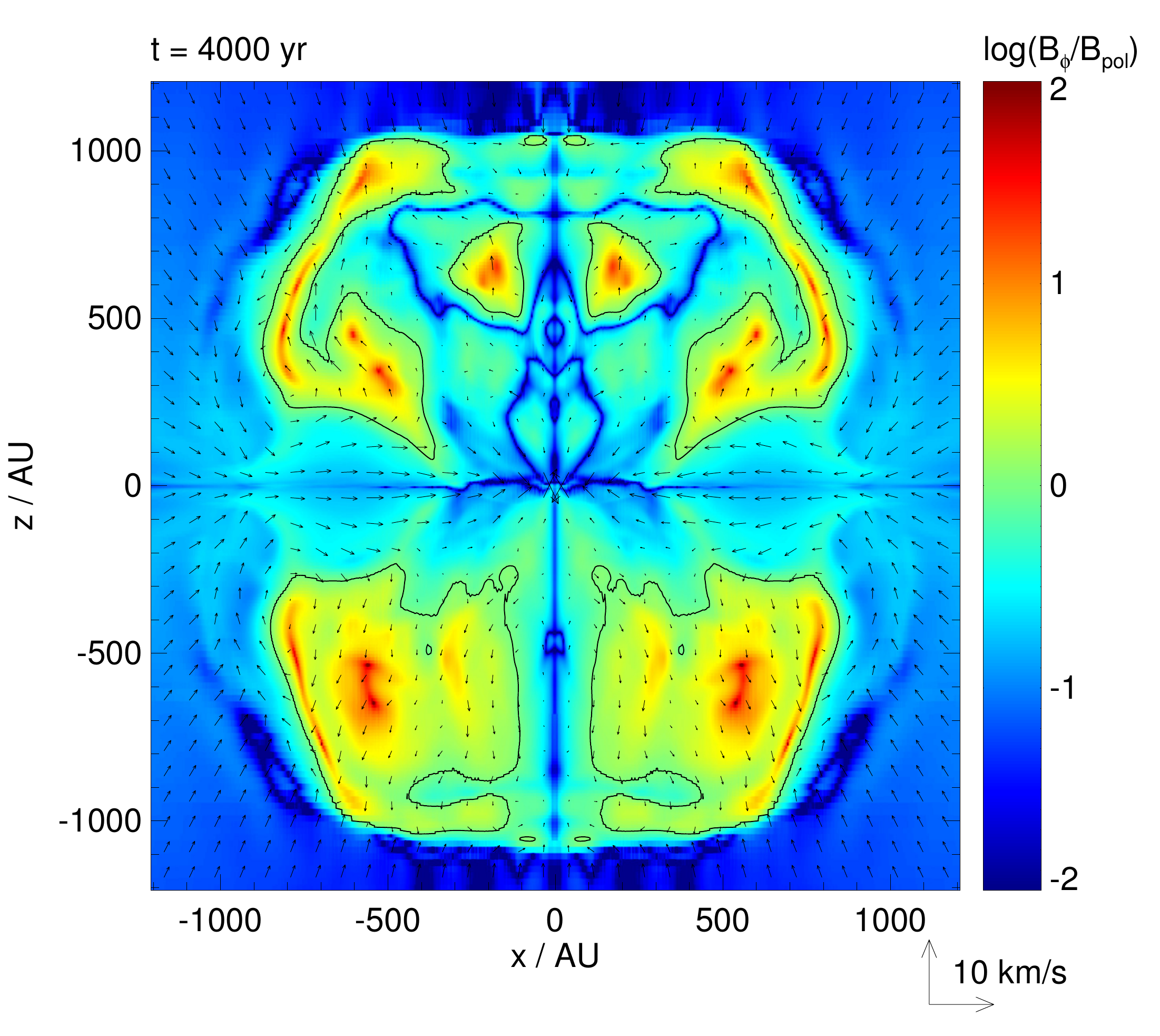} \vspace{-5mm} \\
 \includegraphics[width=84mm]{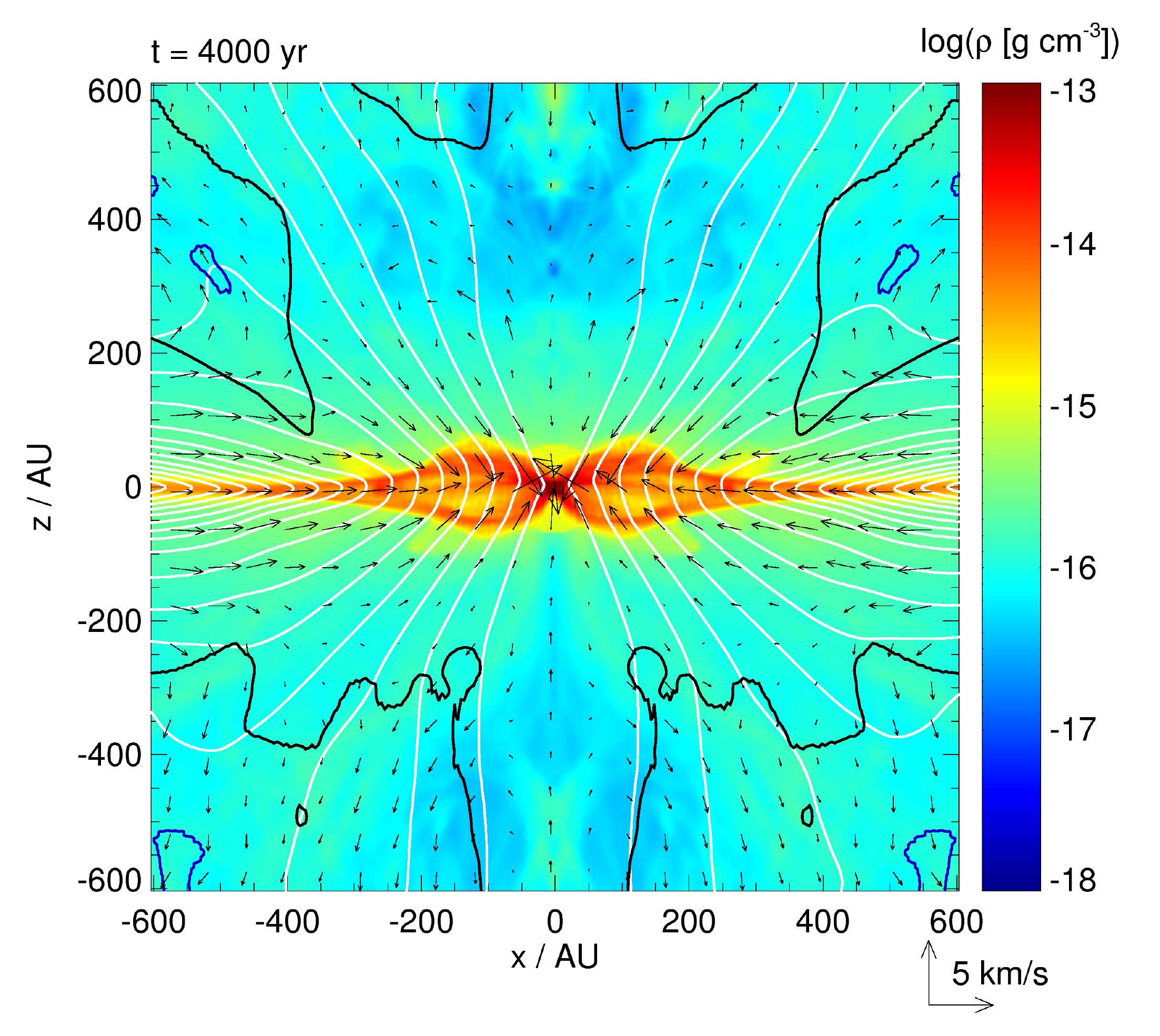}
 \caption{Same as in Fig.~\ref{fig:mag26-4} but for the strong-field run 5.2-4. Top: Additionally the poloidal velocity field (black vectors) is shown. Only parts of the outflow bubble, mainly associated with outflowing gas, are dominated by $B_{\phi}$. Bottom: The strong inclination of the poloidal magnetic field lines is caused by the inwards drag of infalling gas in the disc.}
 \label{fig:mag_52-4}
\end{figure}

The different morphologies of the outflows in run 5.2-4 and run 26-4 raise the question whether the underlying launching mechanisms differ. Firstly, we examine the relative importance of the toroidal and poloidal magnetic field components in the top panel Fig.~\ref{fig:mag_52-4}. Here again -- as in Section~\ref{sec:launch26} -- all quantities like velocity, magnetic field and density are averaged azimuthally before being plotted in order to smooth out local variations which might complicate the analysis. 
As can be seen, a great part of the outflow bubble is not dominated by the $B_{\phi}$ in contrast to the outflow in run 26-4 (see Fig.~\ref{fig:mag26-4}). Interestingly, the regions where $B_{\phi}$ dominates are usually associated with relatively fast outflowing gas. In contrast, the regions which are dominated by $B_{\rmn{pol}}$ either have only slow outflow velocities ($\la$ 1 km s$^{-1}$) or even show infall motions. The fact that outflowing gas seems to be associated with a strong toroidal magnetic field, i.e. $B_{\phi}/B_{\rmn{pol}} > 1$, suggests that the outflow might be driven by the pressure gradient of $B_{\phi}$.

In the bottom panel of Fig.~\ref{fig:mag_52-4} the magnetic field line structure in the inner region is considered in more detail. The poloidal field lines have a strongly pronounced hourglass-shaped configuration with a strong radial component. This is caused by the rapid gas infall in the disc which continuously drags the field lines inwards. Although the poloidal field line structure seems suitable for centrifugal acceleration, i.e. the magnetic field lines are inclined by more than 30$^\circ$ w.r.t. the $z$-axis, only in parts of the region outflowing gas is present, in particular in regions more than 200 AU above/below the disc. Hence, the 30$^\circ$ criterion does not work in this case. 

The failure of the \citeauthor{Blandford82} criterion is not surprising as it was derived for Keplerian disc rotation only and therefore does not apply for strongly sub-Keplerian discs as it is the case here. 
The physical reason for centrifugal acceleration to fail in the inner region are the slow rotation velocities. As therefore the centrifugal force F$_{\rmn{c}}$ is reduced significantly, close to the protostar and the disc F$_{\rmn{c}}$ cannot overcome gravity and consequently gas cannot get flung outwards along the poloidal magnetic field lines. Only at larger radii, where the gravitational force is sufficiently reduced, does centrifugal acceleration work. This fits with the observation that the bulk of outflowing material indeed emerges from radii $\ga$ 300 AU (see Fig.~\ref{fig:mag_52-4}).

\begin{figure*}
 \includegraphics[width=72mm]{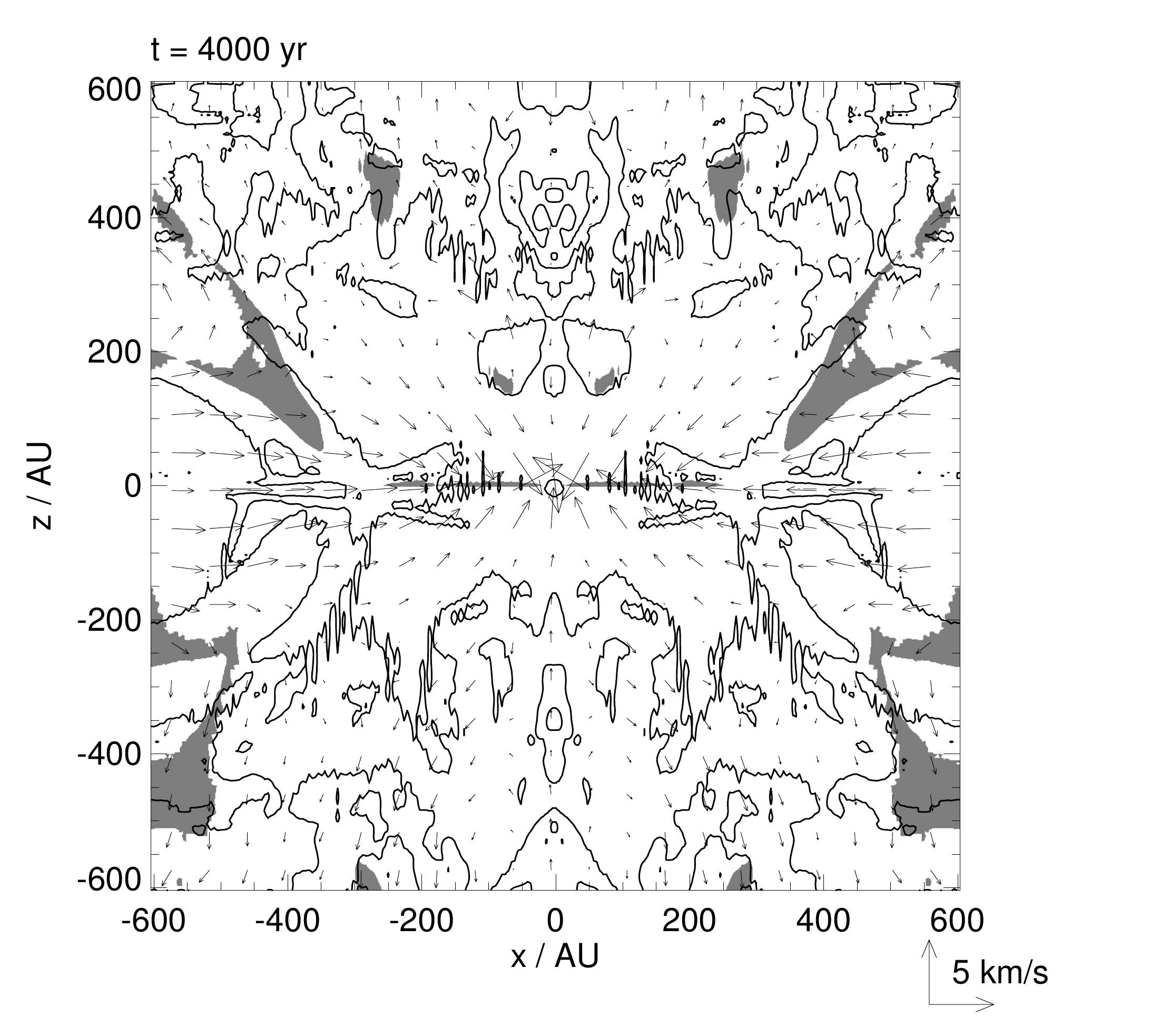}
 \includegraphics[width=72mm]{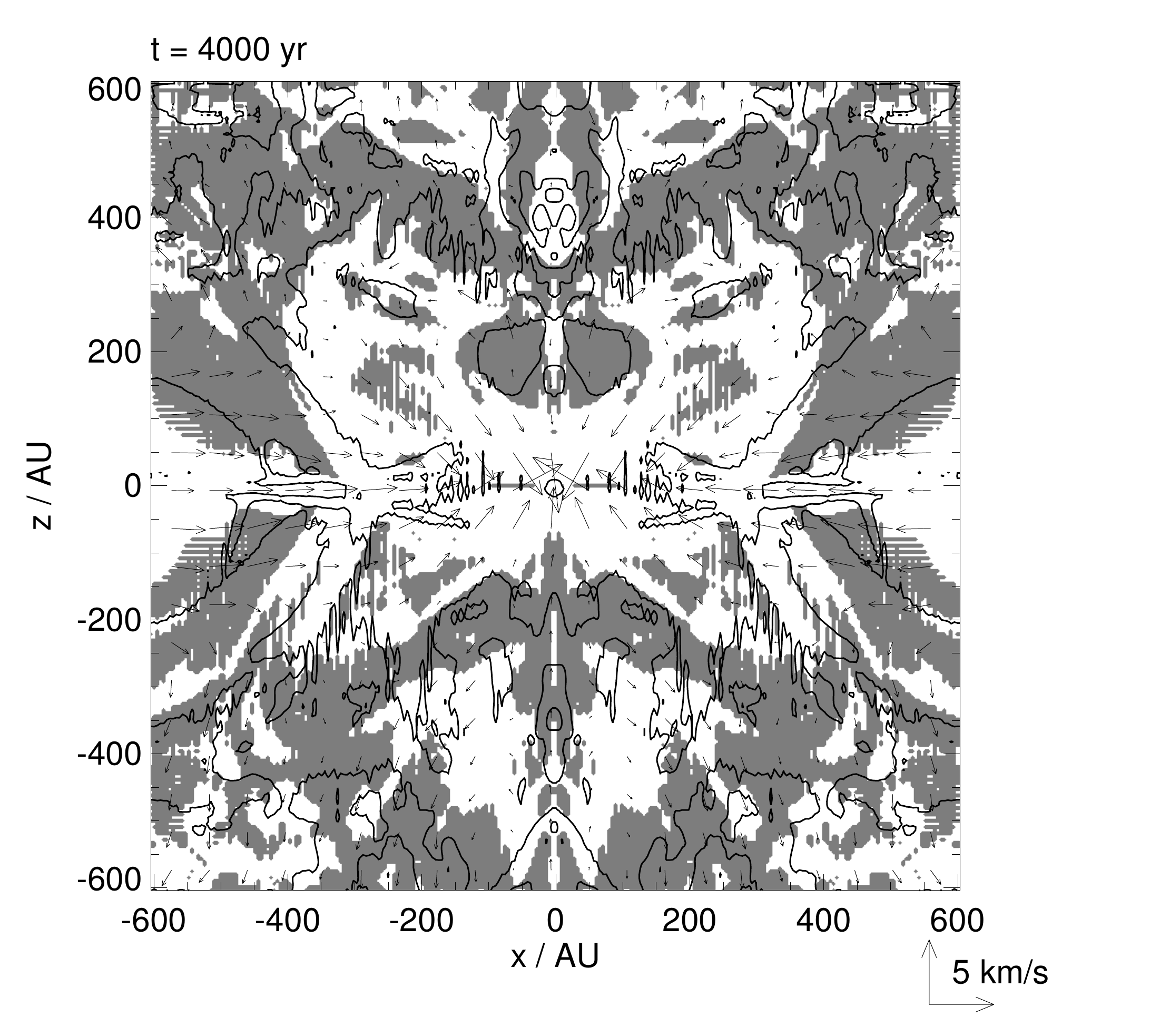}
 \caption{Same as in Fig.~\ref{fig:BP82_26} but for the strong-field run 5.2-4 and after 4000 yr. The black contours enclose the regions where gas feels a real acceleration. Left: Centrifugal acceleration is possible only in a small part of the outflow. Right: The general outflow criterion fits the regions of gas acceleration very well.}
 \label{fig:BP82_52-4}
\end{figure*}

Our centrifugal launching criterion (equation~(\ref{eq:crit})) does not require an underlying Keplerian disc. Hence, we can apply it here as well. We show the result in the left panel of Fig.~\ref{fig:BP82_52-4} whereas the right panel shows the result of applying the general outflow criterion (equation~(\ref{eq:crit2})).
For comparative purposes we also show the regions where gas feels a real, radial outward directed acceleration (black lines)
\begin{equation}
 a_\rmn{r} = \frac{d \mathbfit{v}}{dt} \cdot \mathbfit{e}_\rmn{r} \simeq \left(\frac{d \mathbfit{v}}{dx_\rmn{i}} \frac{dx_\rmn{i}}{dt}\right) \cdot \mathbfit{e}_\rmn{r} = \left(\frac{d \mathbfit{v}}{dx_\rmn{i}} v_\rmn{i} \right) \cdot \mathbfit{e}_\rmn{r} > 0 \, ,
\end{equation}
\begin{figure*}
 \includegraphics[width=80mm]{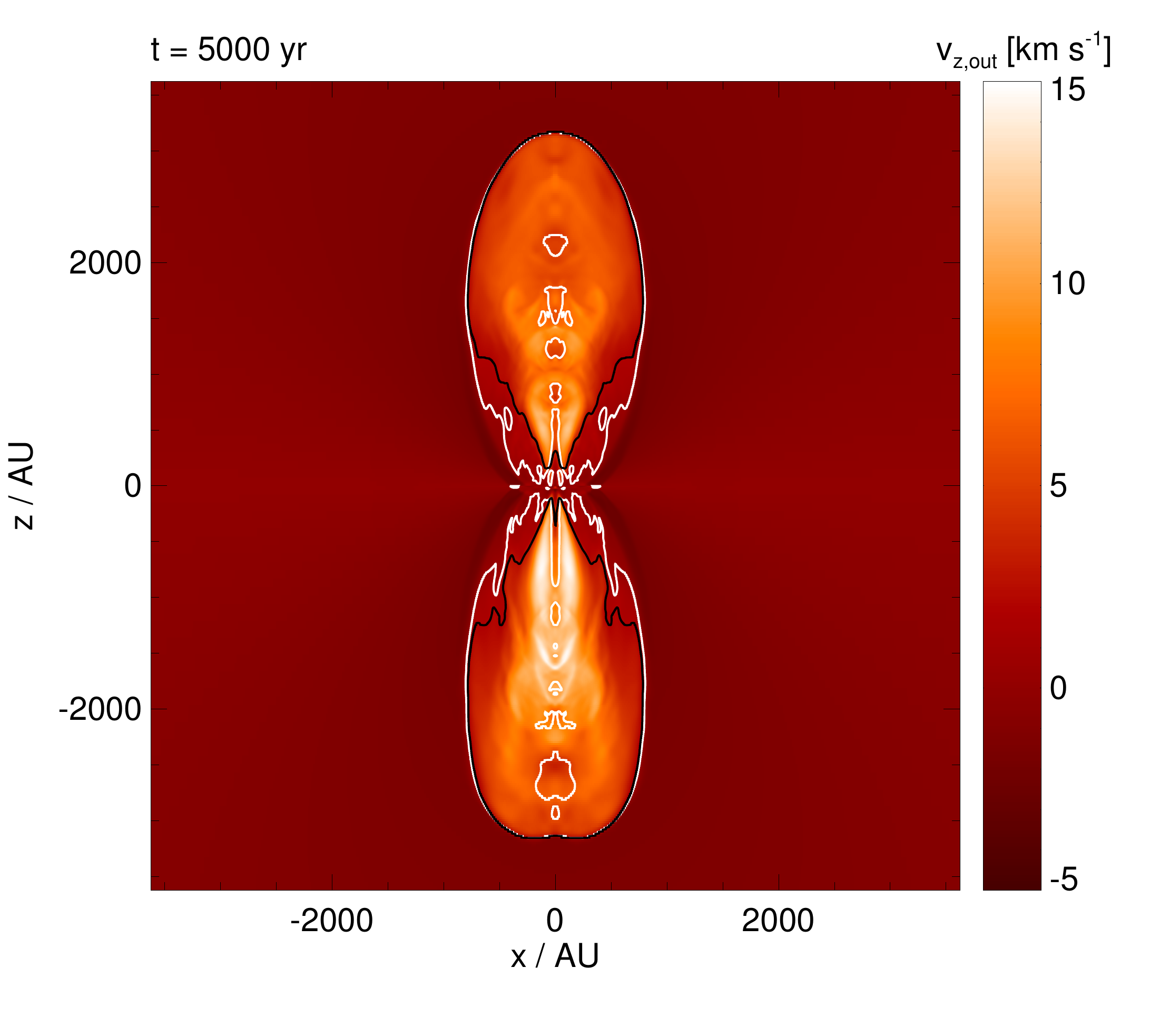}
 \includegraphics[width=80mm]{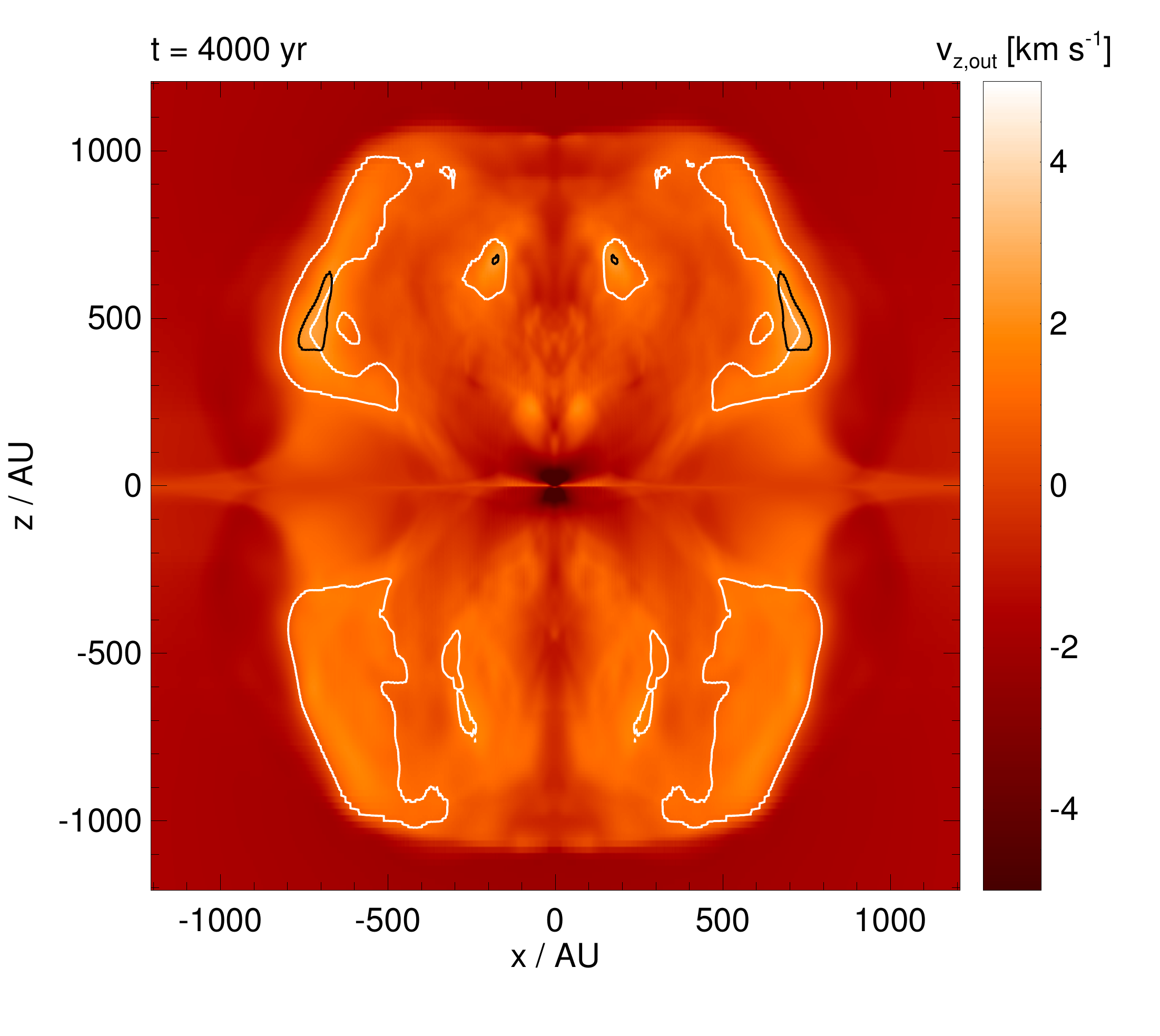}
 \caption{Slice along the $z$-axis for the weak-field run 26-4 after 5000 yr (left) and the strong-field run 5.2-4 after 4000 yr (right). The outflow velocity $v_{z,\rmn{out}}$ is overlaid by the contours where $v_{z,\rmn{out}}$ equals the escape velocity $v_{\rmn{esc}}$ (black line) and the poloidal fast magnetosonic velocity (white line). For the weak magnetic field case 26-4 (left), $v_{z,\rmn{out}}$ is almost everywhere higher than $v_{\rmn{esc}}$ and $v_{\rmn{fast}}$. In contrast, for run 5.2-4 in the most parts $v_{z,\rmn{out}}$ smaller than $v_{\rmn{esc}}$, and exceeds $v_{\rmn{fast}}$ only in parts of the outflow. Note the three times larger spatial scale in the left panel and the different colour scaling.}
 \label{fig:vel}
\end{figure*}
where the time derivative is approximated by spatial derivatives. As already observed in Fig.~\ref{fig:mag_52-4}, regions of acceleration can be found at radii $\ga$ 300 AU but also closer to the $z$-axis and significantly above/below the disc.  Purely centrifugal acceleration, however, only works in the outer regions (left panel of Fig.~\ref{fig:BP82_52-4}). In the upper outflow lobe the grey shaded region fits the outer region of acceleration reasonably well. Although in the lower outflow lobe the agreement is rather poor, it shows that in case of sub-Keplerian disc rotation the criterion clearly works better than the 30$^\circ$ criterion which would predict outflowing gas everywhere. Furthermore, the result suggest that the bulk of outflowing material, which emerges from radii $\ga$ 300 AU (compare upper panel of Fig.~\ref{fig:mag_52-4}), is launched centrifugally also in this case.

However, in particular in the lower outflow lobe and close the $z$-axis, the predicted regions of purely centrifugal acceleration (grey shaded areas in the left panel) hardly match the regions of real acceleration (enclosed by black contours). Hence, as in the weak field case 26-4, we suppose that in these regions the toroidal magnetic field strongly influences the outflow dynamics. Indeed, analysing the outflow with the general criterion given in equation~(\ref{eq:crit2}) taking into account $B_\phi$ (right panel of Fig.~\ref{fig:BP82_52-4}) shows that now the grey shaded regions fit the regions of real acceleration much better, in particular close to the $z$-axis. Naturally, the agreement is not perfect which is probably due to the azimuthal averaging process and the fact the outflow is not stationary as assumed for the derivation of the criterion. Nevertheless, it agrees remarkably well, demonstrating the importance of $B_\phi$. Hence, we tentatively suggest that in this run its influence on the driving is even larger than in run 26-4 where the fast gas is mainly driven centrifugally.

Next, we consider the reason for the poor collimation of the outflow. As mentioned before, outflow collimation is due to the hoop stress produced by $B_{\phi}$. Furthermore, we again note that the disc has strongly sub-Keplerian rotation due to very efficient initial magnetic braking~\citepalias[see][]{Seifried11}. As magnetic braking basically transfers a part of the rotational energy of the disc into energy of the toroidal magnetic field, one could naively expect $B_{\phi}$ in the outflow to be relatively strong and hence the outflow to be well collimated. However, as already indicated in Fig.~\ref{fig:mag_52-4}, $B_{\phi}$ is larger than $B_{\rmn{pol}}$ only in parts of the outflow suggesting a rather moderate toroidal field strength. Indeed, comparing the absolute energy content E$_{\rmn{mag},\phi}$ stored in the toroidal magnetic field of the outflow in the strong-field run 5.2-4 with the weak-field run 26-4 after 4000 yr, reveals that E$_{\rmn{mag},\phi}$ is more than 3 times larger in the latter. The reason for this is that a significant fraction of the angular momentum is removed by magnetic braking already \textit{before} the gas falls onto the disc. Indeed, the disc in run 5.2-4 has a 2 times lower specific angular momentum
\begin{equation}
 l_{\rmn{disc}} = \frac{L_{\rmn{disc}}}{M_{\rmn{disc}}}
\end{equation}
than the disc in run 26-4. Therefore, there is less rotational energy left to transfer into E$_{\rmn{mag},\phi}$. Consequently, $B_{\phi}$ in the outflow and therefore also the hoop stress responsible for collimation are significantly weaker. As at the same time the poloidal magnetic field is quite strong in run 5.2-4, collimation gets even more difficult thus resulting in the poorly collimated, sphere-like outflow.

\citet{Peters11}, studying the interplay of ionising radiation and magnetic fields, find a similar sphere-like outflow structure existing over a timescale of several 10$^4$ yr. The authors simulated a 1000 M$_{\sun}$ core with a resolution of 98 AU and a weaker initial magnetic field ($\mu = 14$) than in our run 5.2-4. Due to the gravitational instability and subsequent fragmentation of the disc as well as the effects of the emerging H~\textsc{ii} regions no well defined Keplerian disc builds up in their simulation. As argued above, this and the resulting weaker $B_{\phi}$ lead to the development of a poorly collimated outflow. Hence, despite different initial conditions the similarities between the outflows observed by \citet{Peters11} and us are not surprising.

To summarise, due to magnetic braking the gas in the disc in run 5.2-4 is rotating relatively slowly. Therefore, $B_{\phi}$ generated by rotation is weak resulting in a poorly collimated outflow.

\subsection{Long term evolution and jet stability} \label{sec:longterm}

Due to computational cost reasons we cannot follow the outflow evolution over more than a few 10$^3$ yr. Nevertheless, we can try to estimate whether these outflows will persist over time or fall back due to the gravitational attraction of the central object and the ram pressure of the infalling gas. For this purpose we compare the outflow velocity $v_{z,\rmn{out}}$ with two basic velocities. The first one to compare with is the escape velocity
\begin{equation}
 v_{\rmn{esc}} = \sqrt{\frac{2 G M_{\rmn{sd}}}{r}}
\end{equation}
where $M_{\rmn{sd}}$ is the mass of the star + disc system. For the outflow to escape the gravitational potential of the central object $v_{z,\rmn{out}}$ has to exceed $v_{\rmn{esc}}$. To take into account the effect of thermal and magnetic pressure in the surrounding gas which also might slow down the outflow, we furthermore compare the outflow velocity to the poloidal component of the fast magnetosonic velocity
\begin{equation}
 v_{\rmn{fast}} \simeq \sqrt{v_{\rmn{A,pol}}^2 + c_{\rmn{s}}^2} \, .
\end{equation}
In Fig.~\ref{fig:vel} we show the azimuthal average of outflow velocity $v_{z,\rmn{out}}$ and the contours where $v_{z,\rmn{out}}$ equals the two velocities mentioned above in a slice along the $z$-axis for run 26-4 (left panel) and run 5.2-4 (right panel).
We first concentrate on run 26-4. As already seen in Fig.~\ref{fig:rprofile}, below 2000 AU the highest velocities occur close to the $z$-axis. Furthermore, it can be seen that the outflow velocity exceeds the escape velocity and also the fast magnetosonic velocity in most parts of the outflow.

\begin{figure}
 \includegraphics[width=72mm]{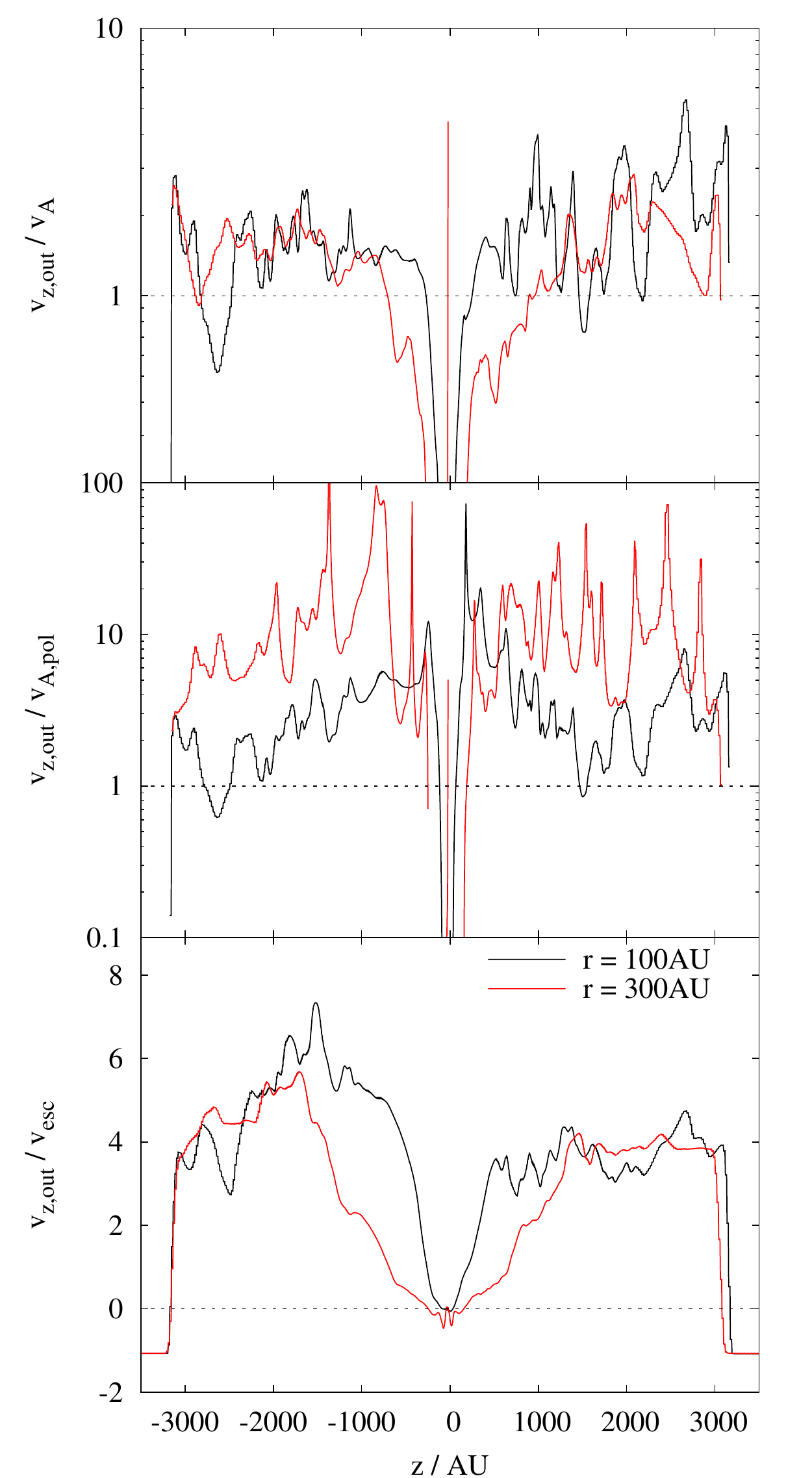}
 \caption{Ratio of $v_{z,\rmn{out}}$ to the Alfv\'enic speed $v_{\rmn{A}}$ (top), to the poloidal Alfv\'enic speed $v_{\rmn{A,pol}}$ (middle) and to the escape speed $v_{\rmn{esc}}$ (bottom) for run 26-4 along a vertical line at a radius of 100 AU (black line) and 300 AU (red line). The outflow velocity is significantly higher than $v_{\rmn{esc}}$ and $v_{\rmn{A,pol}}$ whereas it is comparable to $v_{\rmn{A}}$, probably indicating a self-regulated outflow speed.}
 \label{fig:velslice}
\end{figure}

In Fig.~\ref{fig:velslice} we plot the ratio of $v_{z,\rmn{out}}$ to $v_{\rmn{esc}}$, $v_{\rmn{A,pol}} = \frac{B_{\rmn{pol}}}{\sqrt{4\pi \rho}}$ and $v_{\rmn{A}}$ (equation~(\ref{eq:alf})) for run 26-4 along a vertical line at a radius of 100 AU and 300 AU. As before the quantities are averaged azimuthally.
\begin{figure}
 \includegraphics[width=84mm]{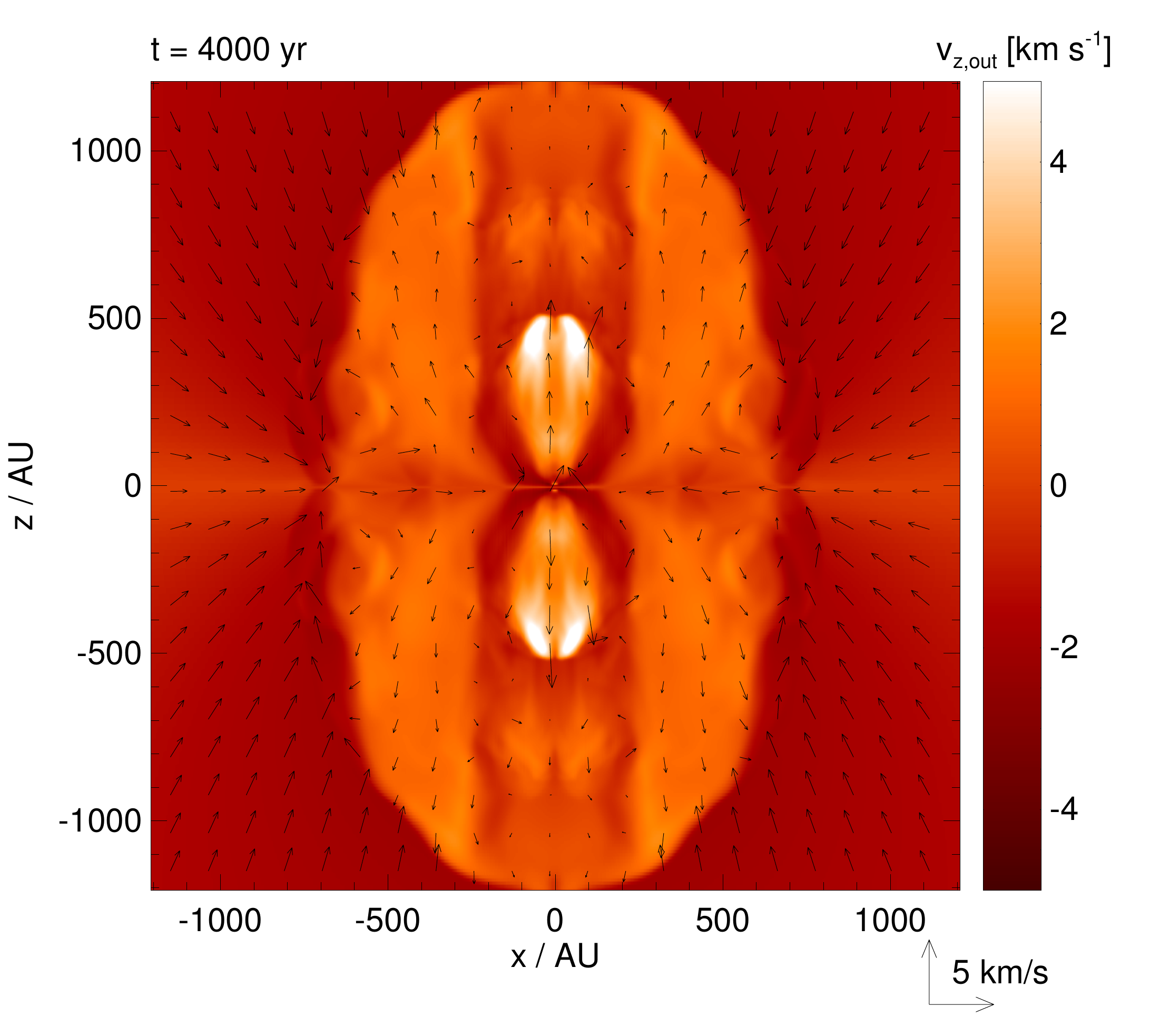}
 \caption{Slice along the $z$-axis in run 10-20 after 4000 yr showing the outflow velocity. Two distinct outflow components are clearly visible, the inner, well collimated, fast jet and the outer, low-velocity outflow.}
 \label{fig:10-20}
\end{figure}
The outflow velocity exceeds the escape speed by up to one order of magnitude (bottom panel of Fig.~\ref{fig:velslice}) and the poloidal Alfv\'enic velocity even by up to two orders of magnitude although the average ratio is 5 - 10 (middle panel of Fig.~\ref{fig:velslice}). We attribute the higher value of $v_{z,\rmn{out}}/v_{\rmn{A,pol}}$ at a radius of 300 AU to somewhat lower densities in this region (compare bottom panel of Fig.~\ref{fig:rprofile}). The absolute outflow velocity, however, decreases with increasing radius as already seen in Fig.~\ref{fig:rprofile} and \ref{fig:vel}. Interestingly, over the whole outflow extension the gas velocity is comparable to or only slightly larger than $v_{\rmn{A}}$. This possibly points to some kind of self-regulation in the outflow where the gas speed is held in the trans-Alfv\'enic range by internal shocks (see Fig.~\ref{fig:26-4}). Once the gas reaches a significantly super-Alfv\'enic speed the flow cannot be stabilised any more leading to instabilities and shocks which in turn reduce the velocity to trans-Alfv\'enic speeds~\citep[see][and upper panel of Fig.~\ref{fig:velslice}]{Ouyed03}. Beyond the shock the gas gets reaccelerated by magneto-centrifugal forces (Section~\ref{sec:launch26}) until it shocks again. Interestingly, the gas speed is comparable to $v_{\rmn{A}}$ and not to $v_{\rmn{A,pol}}$, i.e. $B_{\phi}$ has to be taken into account. This indicates that $B_{\rmn{pol}}$ is not the main agent stabilising the flow although it certainly contributes~\citep{Ray81}. Indeed, \citet{Appl92} show that jets with a toroidal magnetic field are even more stable than jets with a purely poloidal field. In our case we therefore suppose that the stability of the jet is significantly enhanced by $B_{\phi}$.  

Despite ongoing fragmentation of the protostellar disc \citepalias[see][for details]{Seifried11}, the driving of the outflow seems not to decline over time. As furthermore $v_{z,\rmn{out}}$ is significantly larger than $v_{\rmn{esc}}$ and $v_{\rmn{fast}}$, it can be expected that the outflow in run 26-4 will escape the gravitational potential of the central star/disc-system and finally will leave the core even when taking into account the ram pressure of the infalling gas.

Considering run 5.2-4 in the right panel of Fig.~\ref{fig:vel} reveals a markedly different situation. Comparing $v_{z,\rmn{out}}$ to $v_{\rmn{fast}}$ shows that the gas is moving super-Alfv\'enic only in parts of the outer wings of the outflow. Furthermore, a comparison with $v_{\rmn{esc}}$ shows that the outflowing gas is moving almost everywhere with velocities below the escape velocity. In general, $v_{z,\rmn{out}}$ exceeds $v_{\rmn{esc}}$ and $v_{\rmn{fast}}$ by a factor of at most 2. Therefore, when taking into account the additional deceleration of the outflow by the ram pressure of the infalling material, it might be possible that the outflow will re-collapse to the star/disc-system.

To summarise, the outflows of the two simulations presented so far do not only differ in their morphology and their kinematics but possibly also in their longer term evolution. The well collimated outflow is likely to overcome the gravitational potential of the central protostar and leave the core whereas the sphere-like, slowly expanding outflow might be only a short-lived, transient feature in the very early phase of massive star formation possibly recollapsing again.

\subsection{The influence of the initial conditions} \label{sec:global}

So far only the outflows in the runs 26-4 and 5.2-4 have been considered in detail. We have chosen these particular runs as they give representative examples for the outflows observed in the other simulations. The remaining outflows reveal qualitative similarities to one of the two outflows presented before. One of the main results of the previous sections is that the different morphologies can be attributed to the different velocity structure in the disc resulting in different radial positions where the outflows are launched and variations in the strength of the hoop stress ($B_{\phi}$) responsible for outflow collimation. To further confirm this, in the following we consider the whole set of simulations and connect the outflow properties to the properties of the protostellar disc.

We start with the remaining weakly magnetised runs. The outflows in the runs 26-20, 26-0.4 and 26-0.04 show similar morphologies as the outflow in run 26-4. They are reasonably well collimated with collimation factors between 2 in run 26-20 and $\sim$ 4.5 in run 26-0.4 and have outflow velocities preferentially parallel to the $z$-axis. Interestingly, in the runs 10-20  and 5.2-20, whose outflows are rather poorly collimated, a fast and well collimated jet close to the $z$-axis develops after about 2500 yr embedded in the slowly expanding, poorly collimated outflow. Similar applies to run 26-20 with the difference that by the end of the simulation the initial, slowly expanding, poorly collimated outflow is overtaken by the faster, more collimated jet. For demonstrative purposes, in Fig.~\ref{fig:10-20} we show the situation in run 10-20 after 4000 yr.
We mention that similar outflows consisting of two distinct components, an outer, slowly expanding and an inner, fast component have recently also been observed in low-mass star formation simulations~\citep{Banerjee06,Machida08,Hennebelle08,Tomida10,Duffin09,Duffin11}.

Analysing the magnetic field line structure in the runs mentioned before reveals situations similar to that in run 26-4. In the inner region close to the symmetry axis of the jet, where the gas gets accelerated, the magnetic field is only weakly dominated by the toroidal field component, i.e. $B_{\phi}/B_{\rmn{pol}} \la 10$. Furthermore, the field lines at the disc surface are inclined by more than 30$^\circ$ w.r.t. to the vertical axis. As the discs rotate with Keplerian velocities just like in run 26-4 or as in run 5.2-20 with at least about half the Keplerian speed, we therefore suppose that the gas gets launched from the discs by centrifugal acceleration. Applying our outflow criterion (equation~(\ref{eq:crit})) for these runs shows that above the disc the region where purely centrifugal acceleration works is not as extended as in run 26-4 (Fig.~\ref{fig:BP82_26}). With the general criterion for magneto-centrifugal acceleration (equation~(\ref{eq:crit2})), however, we can fit the outflow regions by far better. Hence, considering the jets as purely centrifugally driven might not be an appropriate choice too far from the disc. The evidence shows that the pressure gradient of the toroidal magnetic field contributes significantly to the gas acceleration although, as mentioned before, the launching from the disc itself is most likely to due to centrifugal acceleration. The application of the two criteria also suggest that the dynamics of the outer, slowly expanding outflows in the runs 26-20, 10-20 and 5.2-20 are mainly determined by $B_\phi$.

All remaining runs show poorly collimated, low velocity outflows with outflowing gas emerging almost radially at radii $\ga$ 100 AU and gas infall close to the $z$-axis. Furthermore, the magnetic field line properties in the centre are similar to run 5.2-4 and the protostellar discs driving the outflows are all clearly sub-Keplerian. Moreover, applying the outflow criteria derived in Section~\ref{sec:crit} gives very similar results as for run 5.2. Therefore we conclude that in these runs the outflow driving mechanism is very similar to that in run 5.2-4 , i.e. gas is launched centrifugally from the disc at large radii ($r \ga 100$ AU) with support by $B_{\phi}$. At larger distances $B_{\phi}$ is mainly responsible for the further acceleration (see Section~\ref{sec:launch52}). To remind the reader, the poor collimation is a consequence of the weak toroidal magnetic field in the outflow and of the large launching radii. The larger launching radii are a result of the fact that close to the centre gravity is too strong to be overcome by the weak centrifugal force.

\begin{figure}
 \includegraphics[width=84mm]{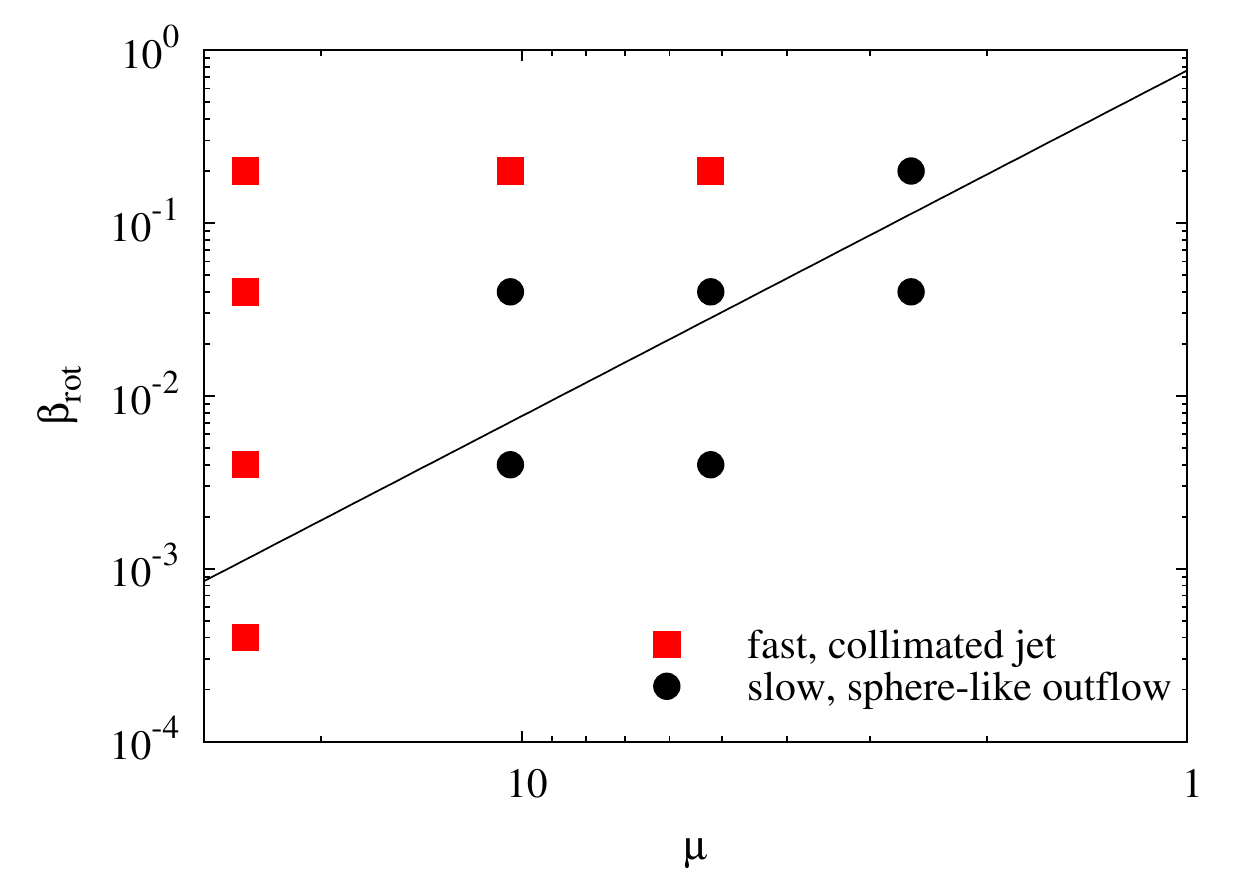}
 \caption{Phase diagram of magnetic field and rotational energy (see Table~\ref{tab:models}) showing the results of the simulations concerning the question of jet formation. For runs with weak magnetic field ($\mu = 26$) or high rotational energies well collimated, centrifugally driven jets develop (red squares). In contrast, for the remaining runs only slowly expanding, poorly collimated outflows are generated (black circles).}
 \label{fig:jets}
\end{figure}

In Fig.~\ref{fig:jets} we show the dependence of the outflow morphology on the initial conditions.
It can be seen that only for the runs with weak magnetic fields ($\mu \ge 26$) or high rotational energies ($\beta_{\rmn{rot}} = 0.20$) do well collimated jets occur. Moreover, all runs with strongly sub-Keplerian discs have poorly collimated outflows~\citepalias[see Fig. 8 in][]{Seifried11}. This clearly demonstrates that for a collimated, high velocity jet to form a disc with (almost) Keplerian rotation is necessary. In contrast, if the disc is rotating strongly sub-Keplerian, both outflow speed and collimation decrease significantly.

Next, we consider the influence of the initial conditions on global outflow properties like mass and momentum. The results for each simulation are listed in Table~\ref{tab:mout}. For the calculation of mass and momentum we only take into account gas more than 47 AU above or below the midplane. By doing so, we avoid the outflow quantities to be affected by material in the protostellar disc. For the momentum we only consider the $z$-component as this is its preferred direction. The outflow masses are of the order of 0.1 - 1 M$_{\sun}$ significantly lower than the sink masses (see also Section~\ref{sec:fluxes}). With momenta of the order of 0.1 - 1 M$_{\sun}$ km s$^{-1}$ we obtain average velocities of a few km s$^{-1}$ lower than typically found in observations~\citep[e.g.][see also Section~\ref{sec:obs} for a detailed discussion]{Beuther02b}.

In general, for a fixed magnetic field strength both centrifugal and $B_\phi$ dominated acceleration predict that the power of the outflow should increase with the rotation speed of the protostellar disc. Thus, for larger amounts of initial rotational energy and hence faster disc rotation, the outflow mass and momentum are expected be higher as well. Indeed, comparing runs with equal $\mu$ in Table~\ref{tab:mout} reveals that except for run 26-20 the outflows are in general more powerful for higher $\beta_{\rmn{rot}}$. We suppose that in run 26-20 the generation of the outflow might be slightly delayed as the material falling onto the disc has a high excess of angular momentum which has to be removed before moving further inwards. Therefore, the accretion rate and consequently also the outflow rates $\dot{M}_{\rmn{out}}$  and $\dot{P}_{\rmn{out}}$ are lower.

Interestingly, there is no clear trend recognizable in $\dot{M}_{\rmn{out}}$ and $\dot{P}_{\rmn{out}}$ when considering simulations with fixed $\beta_{\rmn{rot}}$ but varying $\mu$. In general, for runs with $\mu \ge$ 5.2 there seem to be only small variations within a factor of about 5 in $\dot{M}_{\rmn{out}}$ and $\dot{P}_{\rmn{out}}$ (see Table~\ref{tab:mout}). We note that we exclude run 26-20 in this consideration as it does not fit in this trend. For strong magnetic fields ($\mu$ = 2.6), however, there seems to be a rapid decline in outflow power. Naively, it could be expected that the power of the outflows would increase with magnetic field strength. This is not the case since for stronger magnetic fields disc rotation gets more and more sub-Keplerian~\citepalias[see][for a detailed discussion]{Seifried11}. This results in a weaker centrifugal acceleration and a weaker toroidal magnetic field responsible for further accelerating the outflow. Therefore, for both driving mechanisms the outflow power is expected to decrease for strongly sub-Keplerian discs in agreement with our observations. However, we emphasise that, as the outflow morphologies differ significantly between the individual simulations, the results of this comparison have to be taken with care.

In summary, we find that the generation of a fast ($v_{z,\rmn{out}} \simeq$ 10 km s$^{-1}$), well collimated jet depends on the build-up of a Keplerian disc. In these jets the gas is most likely launched centrifugally from the disc. Somewhat above/below the disc, however, we expect the Lorentz force to come into play contributing to the acceleration of the gas. In contrast, for sub-Keplerian discs slowly expanding, sphere-like outflows develop in which the gas is launched centrifugally from the disc as well although at larger radii. Also here $B_{\phi}$ quickly plays an important role in accelerating the outflow further. Although the morphologies are very different, the global properties like mass and momentum vary only within a factor of 5 showing a decrease towards high magnetic field strengths.

\subsection{The ejection/accretion ratio} \label{sec:fluxes}

Next, we compare the mass outflow rates $\dot{M}_{\rmn{out}}$ with the mass infall rates $\dot{M}_{\rmn{in}}$. To begin with, we limit our consideration to the fiducial runs 26-4 and 5.2-4.
\begin{figure*}
 \includegraphics[width=80mm]{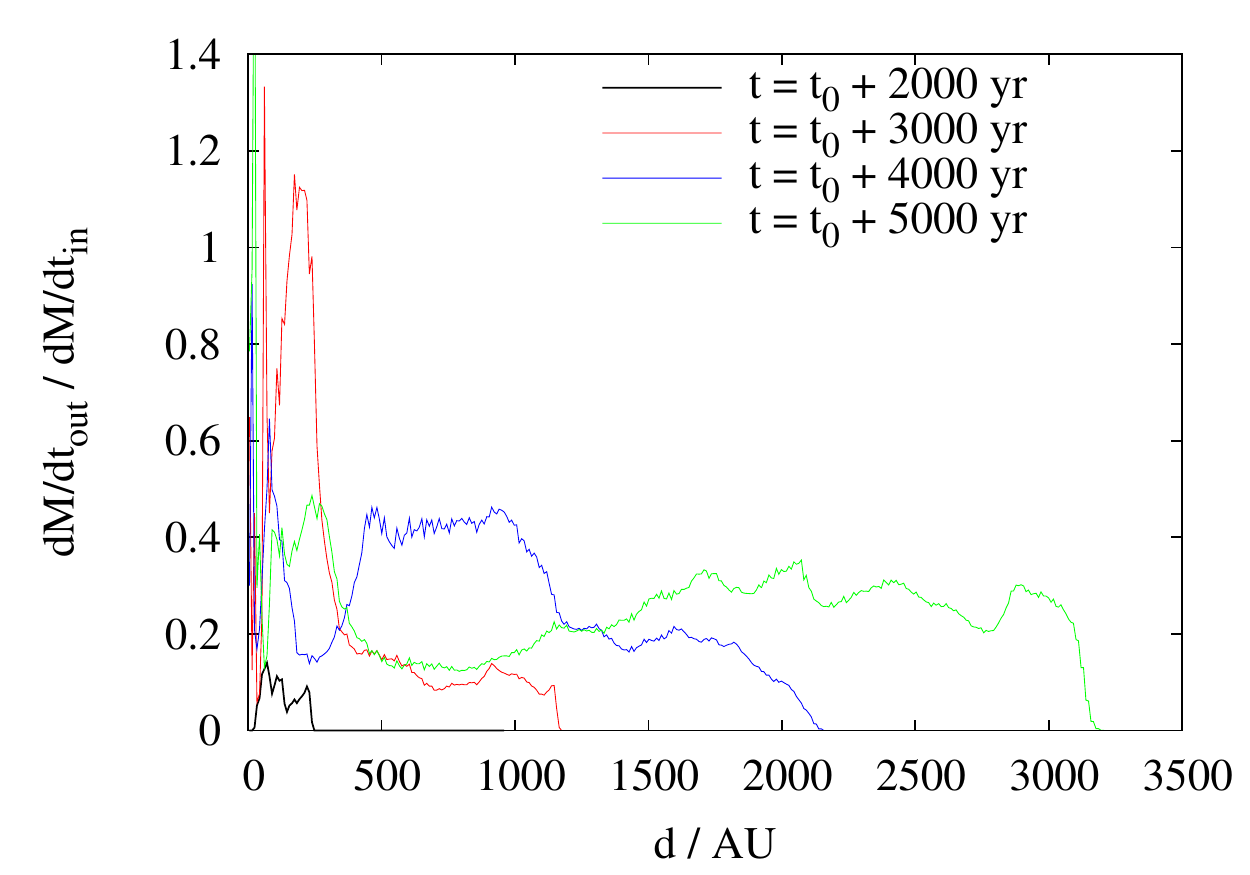}
 \includegraphics[width=80mm]{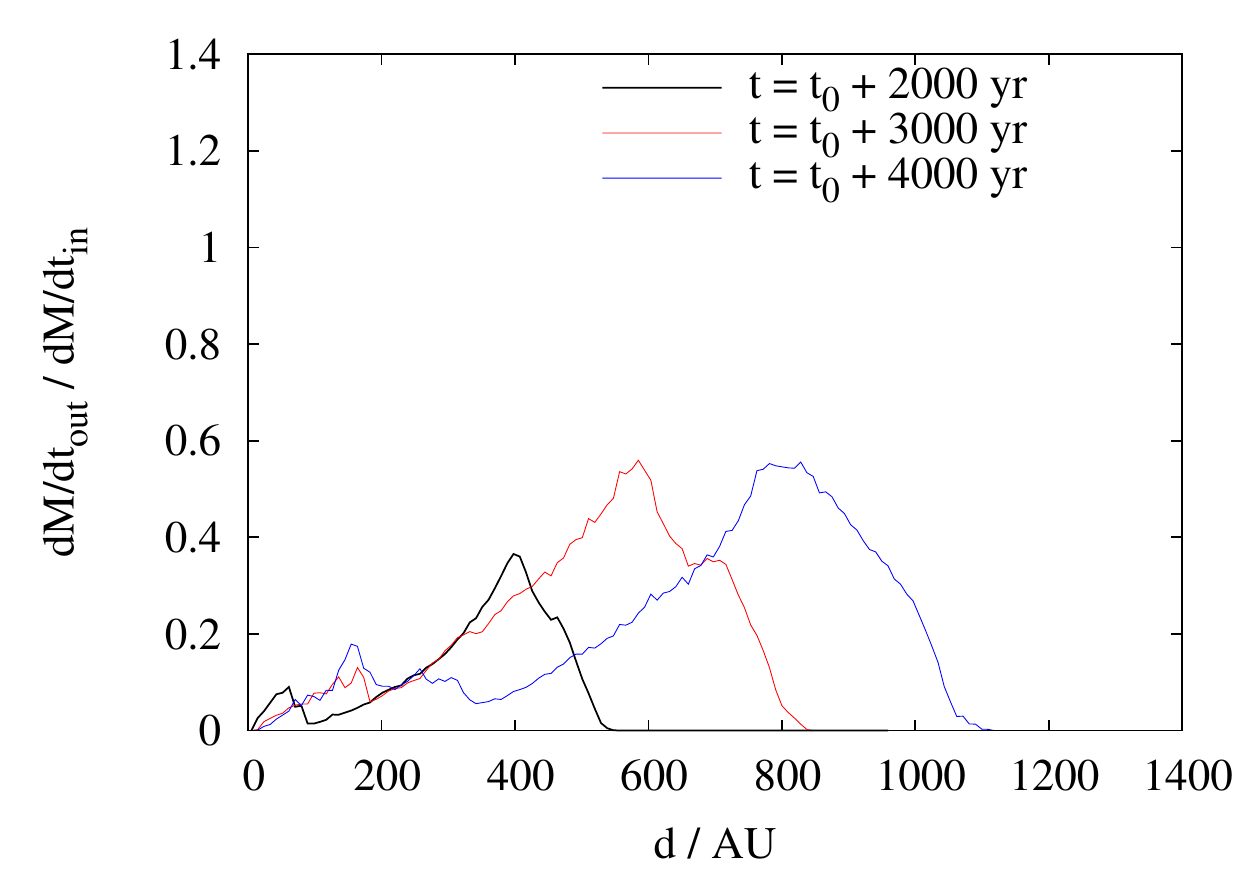}
 \caption{Ratio of the mass outflow rate over infall rate in the weak-field run 26-4 (left) and the strong-field run 5.2-4 (right) as a function of the distance from the centre. The values in both runs agree with theoretical estimates and other numerical work.}
 \label{fig:mflux}
\end{figure*}
In Fig.~\ref{fig:mflux} we show the ratio of outflow to infall rate $\dot{M}_{\rmn{out}}/\dot{M}_{\rmn{in}}$ as a function of the distance $d$ from the centre for different times. The outflow and infall rates are computed on the surface of a cylinder of radius $d$ and height $2 d$ centred around the midpoint of the simulation domain\footnote{We have chosen a cylinder instead of a sphere to lower numerical errors when computing the flux through the surface.}. In run 26-4 $\dot{M}_{\rmn{out}}/\dot{M}_{\rmn{in}}$ is not constant over the outflow extension but varies between 0.1 and 0.4 for $d > 500$ AU. We attribute these variations to the internal shocks occurring in the outflow (compare Fig.~\ref{fig:26-4} and \ref{fig:vel}). Although being somewhat higher, in general the observed range roughly agrees with theoretical estimates of $\sim$ 0.1 obtained for the centrifugal acceleration mechanism~\citep{Pudritz86,Pelletier92}. The high value of $\dot{M}_{\rmn{out}}/\dot{M}_{\rmn{in}}$ at $d <$ 500 AU is due to the circulation flow in the outer parts of the outflow (r $\simeq$ 200 - 300 AU, see bottom panel of Fig.~\ref{fig:mag26-4}). As the outflow speed of this material is not sufficient to escape the gravitational potential, it falls back to the disc thus not contributing to the outflow rate at larger distances. The shape of $\dot{M}_{\rmn{out}}/\dot{M}_{\rmn{in}}$ in run 5.2-4 is qualitatively very different from that in run 26-4. The ratio seems to saturate around a value of $\sim$ 0.1 for late times and small distances ($d \leq$ 400 AU) in agreement with theoretical estimates. For larger distances, however, $\dot{M}_{\rmn{out}}/\dot{M}_{\rmn{in}}$ increases and reaches values up to 0.6.  However, also for this run the measured $\dot{M}_{\rmn{out}}/\dot{M}_{\rmn{in}}$ is in reasonable agreement with theoretical predictions for centrifugal acceleration even though partly somewhat too high.

\begin{figure}
 \includegraphics[width=84mm]{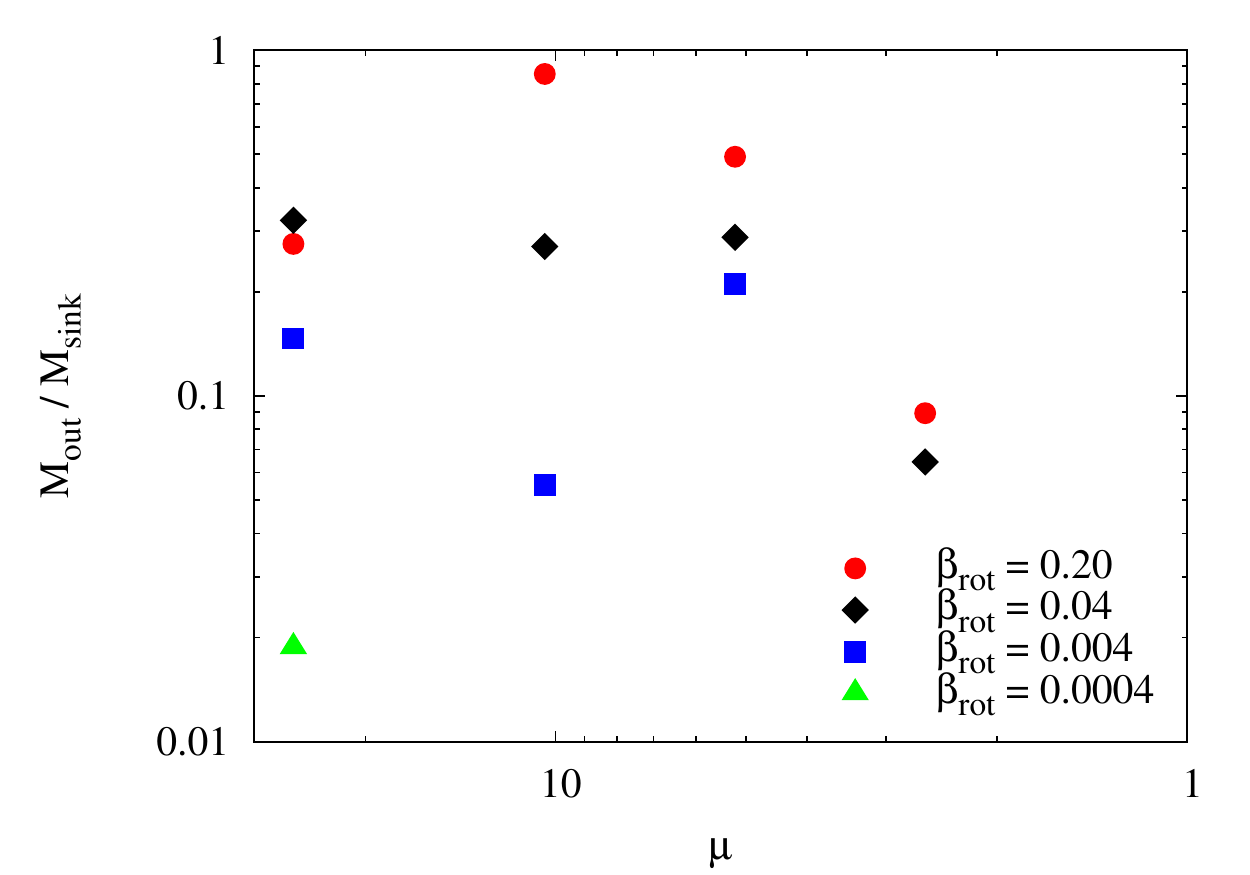}
 \caption{Ratio of outflow mass to sink mass at the end of each simulation as a function of $\mu$ (see Table~\ref{tab:models}). Equal symbols denote equal initial rotational energies. The calculated values scatter around a mean of $\sim$ 0.3 in rough agreement with theoretical estimates.}
 \label{fig:massratio}
\end{figure}

Next, we consider $\dot{M}_{\rmn{out}}/\dot{M}_{\rmn{in}}$ for the whole set of simulations performed. To avoid the problem that even for an individual outflow this ratio is not constant in time and position, we follow a simpler approach than before. For this purpose we approximate $\dot{M}_{\rmn{out}}/\dot{M}_{\rmn{in}}$ by calculating the ratio of the total outflow mass to the total mass accreted onto the sink particles (see Table~\ref{tab:mout}). The result is shown in Fig.~\ref{fig:massratio}.
Excluding one exemption (run 10-20) the ratios are all smaller than 0.5. The values have a mean of $\sim$ 0.3 which agrees very well with other numerical work~\citep{Tomisaka98,Tomisaka02,Hennebelle08,Duffin11} and is only slightly higher than analytical estimates of $\sim 0.1$~\citep{Pudritz86,Pelletier92}. However, the scatter of the individual values is significant ranging from about 0.02 in run 26-0.04 up to 0.9 in run 10-20. Interestingly, run 26-20 has a value of $\dot{M}_{\rmn{out}}/\dot{M}_{\rmn{in}}$ around 0.3 in accordance with the other runs although the absolute values in this simulation are very small (see Table~\ref{tab:mout}). This confirms the explanation given in Section~\ref{sec:global} that in run 26-20 the whole outflow ejection process is slowed down due to the high excess of angular momentum which causes lower accretion rates and in turn also lower outflow rates. Moreover, $\dot{M}_{\rmn{out}}/\dot{M}_{\rmn{in}}$ is generally higher for higher $\beta_{\rmn{rot}}$. As discussed in Section~\ref{sec:global}, this is a consequence of the fact that the outflow power increases with $\beta_{\rmn{rot}}$ whereas at the same time the accretion rates decrease due to an enhanced centrifugal support. For varying $\mu$, however, there is no clear trend in $\dot{M}_{\rmn{out}}/\dot{M}_{\rmn{in}}$ recognizable except a possible drop-off for the highly magnetised runs with $\mu$ = 2.6 which was already observed in Section~\ref{sec:global} for the absolute outflow quantities.

Although the derived values of $\dot{M}_{\rmn{out}}/\dot{M}_{\rmn{in}}$ are in general somewhat higher than analytical predictions, they agree reasonably well with observational results ranging from about 0.1 to 0.3~\citep{Richer00,Beuther02b,Klaassen11}. Furthermore, a ratio of $\sim$ 0.3 is often used to estimate accretion rates of massive protostar from observed outflow mass rates~\citep{Beuther02a,Beuther02b,Beuther03,Beuther04,Zhang05}. As this value fits reasonably well over a wide range of initial conditions, we suppose that accretion rates derived that way seem to be reliable.

\section{Discussion} \label{sec:dis}

\subsection{Launching mechanisms}

Our analysis presented in the Sections~\ref{sec:launch26} and \ref{sec:launch52} suggests that in all simulations the gas is launched from the disc by centrifugal acceleration~\citepalias{Blandford82}. At larger distances above/below the disc the pressure of the toroidal magnetic field significantly contributes to the acceleration of the gas. Although \citet{Lynden-Bell96,Lynden-Bell03} describe outflows whose dynamics are determined by the toroidal magnetic field, we emphasise that our outflows are for the most parts (in time and space) not such so-called magnetic tower flows. In fact, the effect of $B_\phi$ is also implicitly contained in the MHD wind theory (equation~(\ref{eq:crit2})). Therefore, it is more appropriate to consider our outflows in the framework of MHD wind theory as it contains both centrifugal acceleration and the magnetic pressure ($B_\phi$) effects. In such a magneto-centrifugally driven wind the gas is ejected from the disc surface and then gets accelerated towards larger heights. In contrast, in a magnetic tower~\citep{Lynden-Bell96,Lynden-Bell03} gas gets accelerated only at the tip of the outflow -- there is no sustained mass loss coming from the disc. Tower flows are transient phases that arise only in the earliest stages of outflows.

For simulations with weak initial magnetic fields or high rotational energies the situation is as follows: shortly after the formation of the first sink particle, a sub-Keplerian disc builds up driving a slow ($\sim$ 1 km s$^{-1}$) outflow whose driving is probably dominated by the magnetic pressure. As soon as a well defined large-scale Keplerian disc develops, in the inner region a fast ($\sim$ 10 km s$^{-1}$), well collimated jet is launched by centrifugal acceleration. At larger heights and radii $\ga$ 200 AU, however, $B_{\phi}$ significantly contributes to the acceleration. An outer, slowly expanding outflow component is still clearly visible at the end of the simulations 26-20, 10-20 and 5.2-20 (see Fig.~\ref{fig:10-20}). Such outflows consisting of two components, an outer, slowly expanding and an inner, fast component have also been observed in simulations of low-mass star formation~\citep{Banerjee06,Machida08,Hennebelle08,Tomida10,Duffin09,Duffin11}.

For runs with strong magnetic fields, the protostellar discs are sub-Keplerian all the time due to strong magnetic braking~\citepalias[see][for a detailed discussion]{Seifried11}. Therefore, during the time covered by our simulations no fast, central jet component develops and only a slowly expanding, poorly collimated bubble is present. The gas seems to be mainly launched centrifugally although the toroidal magnetic field has probably a large impact on driving the outflow already close to the disc. We tried to clarify the situation by applying a criterion derived in Section~\ref{sec:crit} depending on the actual position, the protostellar mass and the local rotation speed (equation~(\ref{eq:crit})). In general, using the criterion for the present simulations indicates that it is applicable despite the constraints like axis-symmetry, stationarity and corotation assumed for its derivation. We show that for strongly sub-Keplerian discs it acceptably fits the region where the outflow gets launched (see left panel of Fig.~\ref{fig:BP82_52-4}) in contrast to the \citeauthor{Blandford82} criterion which fails when analysing the sub-Keplerian cases. Together with the results from run 26-4 this demonstrates that simply considering the inclination of the magnetic field w.r.t. the $z$-axis might not be sufficient to determine the driving mechanism nor the region where the outflow gets launched. On the one hand even for inclination angles smaller than 30$^\circ$ the outflow can be driven centrifugally (see Fig.~\ref{fig:mag26-4} and \ref{fig:BP82_26}) whereas on the other hand there are situations where purely centrifugal acceleration does not work despite inclination angles larger than 30$^\circ$ (see Fig.~\ref{fig:mag_52-4} and \ref{fig:BP82_52-4}).

Again, we emphasise that in all runs the mechanism responsible for outflow launching is centrifugal acceleration. Only at larger heights do the effects of the toroidal magnetic field come into play. The situation is therefore not such that the outflow is either driven centrifugally or by the toroidal magnetic pressure. Such a classification often found in literature is an oversimplification not describing the situation appropriately as both processes play an important role. We have shown that there is a continuous transition, from centrifugal dominated acceleration to acceleration dominated by $B_\phi$. These two regimes are both implied in the solution of the stationary MHD equations presented in Section~\ref{sec:crit}~\citep[see also Section 4.1 of][]{Spruit96}. The different outflow morphologies mainly result from the varying strength of $B_{\phi}$ responsible for collimation as for strongly sub-Keplerian cases $B_{\phi}$ is too weak to properly collimate the outflows.

As mentioned before, several authors have simulated outflows consisting of two different components. However, their conclusions as which component is driven by which mechanism partly contradict each other. On the one hand,
several authors~\citep{Banerjee06,Duffin09} claim that the outer, larger scale outflow is driven by the magnetic pressure~\citep{Lynden-Bell03} whereas the inner, fast outflow embedded in the first one is considered as a centrifugally driven jet~\citepalias{Blandford82}. On the other hand, \citet{Machida08} and \citet{Tomida10} find the situation to be exactly vice versa. All authors argue on the basis of an analysis of the field line structure such as the ratio of $B_\phi$ to $B_{\rmn{pol}}$ or the field line inclination w.r.t. the protostellar disc. As we argued above, this kind of consideration alone can be misleading which might have caused the contradictory results. Hence, it would be interesting to apply the criterion derived in this work to confirm the results in an independent way. This was recently done by~\citet{Duffin11} in simulations of low-mass cores. The results support our interpretation of a magnetic pressure driven, outer outflow component and a centrifugally driven, inner component. Furthermore, this demonstrates the applicability of our criterion in equal measure to low-mass and high-mass protostellar outflows without any restrictions.

\citet{Hennebelle08} studying the influence of the initial magnetic field strength on the formation of outflows around low-mass stars find striking similarities to our results, i.e. a well collimated, fast outflow component embedded in a slowly expanding component for weak magnetic fields and a poorly collimated outflow for strong fields. Their interpretation of a centrifugally driven wind for the latter case and a magnetic tower flow for the former case, however, is in opposition to our interpretation. Hence, it would be interesting to apply our criterion derived here to confirm this conclusion in another way. Differences might also result from the fact that \citet{Hennebelle08} consider a low-mass core which is less gravitationally unstable. Therefore, in particular in the strongly magnetised case the situation might be less sub-Keplerian possibly leading to different results.

In general, our results agree remarkable well with those of low-mass star formation simulations even over the wide range of initial conditions covered in out work. However, the good agreement between our and the low-mass case results could not be expected a priori as in contrast to low-mass cores the cores presented here are highly gravitationally unstable (about 56 Jeans masses). Therefore the dynamics and timescales of the whole system can differ significantly from low-mass cores. Another difference, though not covered in this work, is the influence of initial turbulence and radiative feedback of the protostars. However, as we are still in a very early evolutionary stage, we suppose that the effect of radiation is still rather limited. In contrast, the influence of turbulence which we will examine in a subsequent paper is expected to be significant~\citep[see also][]{Hennebelle11}.

\subsection{The evolution and impact of outflows}

For typical massive star forming regions the observed mass-to-flux ratios are usually only slightly supercritical, $\mu \la 5$~\citep{Falgarone08,Girart09,Beuther10}. Based on the results obtained in this work, the question arises how well collimated, fast outflows frequently observed around high-mass protostellar objects can form~\citep[e.g.][]{Beuther02b,Beuther04,Beltran11}. As we showed that the formation of such outflows is linked to the build-up of Keplerian discs, the question reduces to how the magnetic braking efficiency can be reduced allowing Keplerian discs to form. A possible solution is provided by the poorly collimated outflows themselves evacuating the region above and below the disc therefore significantly reducing the magnetic braking efficiency~\citep{Mouschovias80}. Furthermore, a misalignment between the rotation axis and the initial magnetic field~\citep{Hennebelle09,Ciardi10} or non-ideal MHD effects like ambipolar diffusion, ohmic dissipation or the Hall effect could reduce the magnetic braking efficiency, in particular in later stages~\citep[e.g.][]{Dapp10,Li11}. All these possibilities lead to a successive build-up of Keplerian discs during the protostellar evolution towards Class I/II objects, hence allowing the generation of well collimated, high-velocity jets~\citepalias[see also][for a more detailed discussion]{Seifried11}.

As shown in Fig.~\ref{fig:zpro26-4}, the position-velocity (PV) diagram shows a typical ''Hubble law``-like behaviour also seen in observations~\citep[e.g.][]{Lada96,Beltran11}. Such an behaviour is a natural outcome of several wind models like the jet-driven bow shock model or the wind-driven shell model~\citep[see][for an overview]{Cabrit97}. However, the comparison between our results and those models is not straightforward as in those wind models the velocity increase with distance is only an apparent and no real acceleration. This apparent acceleration is a consequence of the inclination of the outflow w.r.t. the plane of the sky and the fact that mainly gas swept up in the bow shock and not inside the outflow lobe contributes to the observable emission. In contrast, in our case the observed acceleration is real and is mainly due to the gas inside the bow shock structure whereas the material in the shock itself has mainly negative velocities (compare Fig.~\ref{fig:rprofile}). Hence, an observed acceleration of the bulk velocity in outflows can also be real and not only due to projection effect.

However, an interesting point made by those wind models is the fact that the outflows partly consist of gas entrained from the ambient medium. We tested this statement by implementing a passive mass scalar in run 26-4 which reveals an outflow morphology similar to that assumed in the wind models. This mass scalar is set to 1 in a disc of 75 AU above and below the midplane and gets advected with the gas. Hence, we can test how much of the gas in the outflow stems from this disc and how much from the ambient medium. The analysis shows that by the end of the simulation about 92\% of the gas mass in the outflow stems from the disc. Hence, most of the gas gets accelerated from the bottom of the outflow up to the tip and only a small part is entrained from the ambient medium. A possible explanation might be the high infall velocities which are a result of the highly gravitationally unstable configuration. We suggest that the gas hitting the bow shock it is mainly deflected sidewards and gets channelled downwards along the shock rather than getting stuck and being entrained with the outflow. The situation might change if the core is less gravitationally unstable and the infall velocities are smaller. We tentatively suppose that in this case it is more likely that the gas gets entrained by the outflow. 

\citet{Yorke02} and \citet{Krumholz05} suppose that outflows create low density regions along the polar direction which aid the formation of massive stars by channelling the radiation outwards, therefore diminishing the radiation pressure and allowing accretion to proceed. As seen in bottom panel of Fig.~\ref{fig:mag26-4}, the material ejected from the disc indeed carves a low density region into the vicinity. It would be interesting to see whether the density decrease in our simulations is sufficient to allow radiation produced by the protostar to escape efficiently as recently seen in radiation-hydrodynamical simulations by~\citet{Cunningham11}. Furthermore it would be of interest to see to what extent the radiation pressure would decollimate the outflow as proposed recently by \citet{Vaidya11} on the basis of 2-dimensional jet simulations.

\subsection{Comparison to observations} \label{sec:obs}

In Table~\ref{tab:mout} we have listed the mass and momentum outflow rates of all runs performed. The calculated mass-loss rates vary between $1.6 \cdot 10^{-5}$ and $2.7 \cdot 10^{-4}$ M$_{\sun}$ yr$^{-1}$. This is in agreement with typically observed mass loss rates in high-mass star forming regions ranging from 10$^{-5}$ up to a few 10$^{-3}$ M$_{\sun}$ yr$^{-1}$ although at the lower end~\citep[e.g.][]{Beuther02b,Zhang05,Wang11,Ren11}. Like our simulations those observations refer to the very early stage of massive star formation although the calculated dynamical timescales are of the order of a few 10$^3$ up to a few 10$^4$ yr, typically somewhat larger than ours. However, as for a large fraction of the observed systems no signs of H~\textsc{ii}-regions have been detected, the outflows are probably still magneto-centrifugally driven winds which is why our mass-loss rates are comparable.

The calculated momentum outflow rates in our models range from $4.0 \cdot 10^{-5}$ to $4.7 \cdot 10^{-4}$ M$_{\sun}$ km s$^{-1}$ yr$^{-1}$. This is at the lower end usually reported by observations of massive protostellar outflows finding $\dot{P}_{\rmn{out}}$ of the order of 10$^{-4}$ to 10$^{-2}$ M$_{\sun}$ km s$^{-1}$ yr$^{-1}$~\citep[e.g.][]{Beuther02b,Zhang05,Shi10,Wang11}. As the outflow velocity in centrifugally driven jets is coupled to the rotation velocity~\citep{Michel69,Pelletier92}, the maximum outflow velocities which can be reached in our simulations, are of the order of 10 km s$^{-1}$ (see equation~(\ref{eq:vrotmax})), about one order of magnitude lower than typically observed. Therefore, the lower values of $\dot{P}_{\rmn{out}}$ in our runs are not surprising. However, we expect our momentum outflow rates to increase and to become comparable to observations if we run the simulations over comparable timescales of a few 10$^4$ yr and with a higher resolution which allows for faster outflow material arising from deeper in the gravitational potential well.

Observational results on outflow morphologies are still somewhat ambiguous. One the one hand, outflows around massive protostars are usually found to be less collimated than their low mass counterparts revealing collimation factors of the order of 1 - 2~\citep[e.g.][]{Ridge01,Wu04}. On the other hand, \citet{Beuther02a,Beuther04} also find highly collimated, massive outflows with collimation factors as high as 10. Based on this, \citet{Beuther05} propose an evolutionary scenario in which well collimated, magnetically driven outflows occur in the very early phase of massive star formation. In their further evolution the outflows get progressively less collimated due to the build-up of H~\textsc{ii}-regions. According to this, early-stage outflows as presented here should all be more or less well collimated. Obviously this is not the case, as only runs with a weak magnetic field ($\mu$ = 26) reveal well collimated outflows with collimation factors up to 4.5. Therefore, our results suggest that the collimation not only depends on the evolutionary stage but also on the initial conditions of the molecular cloud core, in particular on the magnetic field strength.

\subsection{The evolution of outflow collimation}

Our results suggest that during the very early stages (10$^3$ - 10$^4$ yr) outflows in typically magnetised, massive cores ($\mu \la 5$) should rather be poorly collimated with collimation factors of 1 - 2 instead of 5 - 10. This agrees with a number of observations of outflows around young massive protostellar objects~\citep[e.g.][]{Ridge01,Torrelles03,Wu04,Sollins04,Surcis11}. Therefore, we suggest that in the earliest stage, i.e. even before the scenario described by \citet{Beuther05} applies, outflows are indeed rather poorly collimated except in case of an unusually weak magnetic field. In their further evolution, however, the collimation will increase quickly due to the development of a fast, central jet coupled to the build-up of a Keplerian disc. Such a behaviour can in particular be seen in the runs 26-20, 10-20 and 5.2-20. Hence we suggest that, beside tracing later stages as proposed by \citet{Beuther05}, poorly collimated outflows around massive protostars could also trace the very early stage of massive star formation. Hence, one might have to be careful when trying to infer the stage of star formation from the collimation of the outflow.

To confirm such an evolutionary scenario as proposed here, observations of the earliest stage of massive star formation would be necessary in combination with magnetic field measurements. Unfortunately, to date such observations are difficult to obtain and therefore rather rare. However, there is an interesting observation which supports the picture of very early stage, poorly collimated outflows successively collimating over time. Observing two spatially adjacent, massive protostars in the star forming region W75N, \citet{Torrelles03} and \citet{Surcis11} find the younger of the two having a spherical outflow whereas the more evolved protostar has a well collimated outflow. Due to the close proximity to each other, the authors expect the environmental conditions to be very similar and hence not to cause the different morphologies. Therefore the authors conclude that the differences are rather a consequence of different evolutionary stages and that the younger, poorly collimated outflow is possibly only a transient feature. This interpretation fits perfectly into our evolutionary scenario in which after an initial phase a poorly collimated outflow gets overtaken by a well collimated jet.

\section{Conclusions} \label{sec:conclusion}

We have studied the collapse of massive molecular cloud cores with varying initial rotational and magnetic energies. The mass-to-flux ratios of the magnetically supercritical cores range from 2.6 up to 26. The cores have rotational energies well below the gravitational energy and contain about 56 Jeans masses. Hence, they are highly gravitationally unstable and possible sites of massive star formation. In this work we focussed on the launching mechanism and the properties of early outflows and jets. Furthermore, based on the stationary, axisymmetric MHD equations we derived a generalised criterion to determine the launching mechanism of the outflows. The criterion is applicable to the whole outflow region and to situations with sub-Keplerian disc rotation. In the following we summarise our main findings.
\begin{enumerate}
 \renewcommand{\theenumi}{\arabic{enumi}.}
 \item We show that our outflow criterion (equations~\ref{eq:crit2} and \ref{eq:crit}) can successfully distinguish between the different driving mechanisms and that it works over the complete extension of the outflows. We discuss an example where the outflow is centrifugally driven up to about 800 AU thereby demonstrating that considering only the ratio of $B_{\phi}$ to $B_{\rmn{pol}}$ is not sufficient to determine the driving mechanism. Furthermore, we successfully apply the criterion to runs with strongly sub-Keplerian protostellar discs where the frequently used 30$^\circ$-criterion for the inclination of magnetic field lines fails.
 \item Depending on the initial magnetic field strength two types of outflows are observed: Well collimated, high-velocity jets for runs with weak magnetic fields ($\mu \ge 26$) and high rotational energies ($\beta_{\rmn{rot}} = 0.20$) and poorly collimated, slowly expanding outflows for runs with strong fields. The development of fast jets is coupled to the build-up of Keplerian discs. In none of the strongly magnetised runs with strongly sub-Keplerian disc rotation a well collimated jet is observed.
 \item In all runs centrifugal acceleration is responsible for launching the gas from the discs. With increasing distance from the disc the pressure gradient of the toroidal magnetic field progressively contributes to the further acceleration of the gas. For runs with sub-Keplerian discs the outflows are centrifugally dominated at large radii ($\ga$ 100 AU) where gravity is sufficiently reduced to be overcome by the centrifugal force. In these outflows $B_\phi$ seems to play a more important role in the driving than in the fast jets.
 \item The morphological differences of the outflows are mainly due to the varying strength of the hoop stress responsible for outflow collimation. For sub-Keplerian discs the generation of $B_{\phi}$ happens more slowly, so that their corresponding outflows are less collimated.
 \item We show that an outflow can be maintained despite the fragmentation of the protostellar disc. Furthermore, knotty outflow structures can also be produced in continuously fed jets by gas repeatedly experiencing shocks and reaccelerating again even far from the disc. For such outflows the toroidal magnetic field seems to contribute significantly to the overall stabilisation.
 \item The observed mass and momentum outflow rates are of the order of 10$^{-4}$ M$_{\sun}$ yr$^{-1}$ and 10$^{-4}$ M$_{\sun}$ km s$^{-1}$ yr$^{-1}$, respectively, in reasonable agreement with observational results. The calculated mass ejection/accretion ratios scatter around a mean of 0.3 in agreement with both theoretical estimates and observational results. 
 \item Based on the results of the strongly magnetised simulations ($\mu \la 5$), we suggest an evolutionary scenario where a poorly collimated outflow is typical for the very early stage of massive star formation. Over time the outflow collimation will increase due to the development of a well collimated, fast jet overtaking the slowly expanding outflow. This picture is also supported by observations. Furthermore, analysing the sphere-like, slowly expanding outflows suggest that they are possibly only transient features which might re-collapse during their further evolution.
\end{enumerate}

We showed that a simple approach such as the value of $B_{\phi}/B_{\rmn{pol}}$ or the inclination of magnetic field lines is not sufficient to determine the launching mechanism of outflows. This could be the reason for the partly conflicting results found in literature. We therefore strongly suggest to use a self-consistent criterion which is applicable to the whole outflow region as well as sub-Keplerian discs frequently found in numerical simulations~\citep[e.g.][]{Mellon08,Hennebelle08,Duffin09,Seifried11}. Moreover, we emphasise that a separation in purely centrifugally or magnetic pressure driven winds is probably an oversimplification not describing the situation appropriately. We showed that magneto-centrifugal acceleration has two regimes, a centrifugally dominated and a $B_\phi$ dominated one and that within a realistic outflow there is a continuous transition from one regime to the other.

A growing number of observations of discs and well collimated, bipolar outflows around high-mass protostars \citep[see][for recent reviews]{Beuther05,Cesaroni07} support a high-mass star formation scenario via disc accretion similar to low-mass star formation. On the other hand, observations also show that prestellar cores with masses ranging from 2 - 2000 M$_{\sun}$ are usually only slightly supercritical with $\mu \la 5$ \citep{Falgarone08,Girart09,Beuther10}. Together with our numerical results this suggest that typically there should be no well-collimated jets in the very early stages of high-mass star formation but rather sphere-like, slowly expanding outflow structures. As discussed in the paper, the question of how well collimated jets are generated breaks down to the problem of how Keplerian discs can be formed in highly magnetised cores. For this to happen, effects in the later evolution of the system are required that reduce the efficiency of magnetic braking. In particular, the initial, slowly expanding outflows as well as non-ideal MHD effects could accomplish, this resulting in a successive growth of discs and the development of well collimated jets over time.

\section*{Acknowledgements}

The authors like to thank the anonymous referee for his comments which helped to significantly improve the paper. The authors also like to thank Christian Fendt, Philipp Girichidis and Christoph Federrath for many helpful discussions and suggestions. The simulations presented here were performed on HLRB2 at the Leibniz Supercomputing Centre in Garching and on JUROPA at the Supercomputing Centre in J\"ulich. The FLASH code was developed partly by the DOE-supported Alliances Center for Astrophysical Thermonuclear Flashes (ASC) at the University of Chicago. D.S. and R.B. acknowledge funding of Emmy-Noether grant BA3706 by the DFG. D.S. especially thanks the Heidelberg Graduate School of Fundamental Physics for funding a research visit at McMaster University, Hamilton. D.D. is supported by McMaster University and R.E.P by a Discovery grant from NSERC of Canada. R.S.K.~acknowledges subsidies from the {\em Baden-W\"urttemberg-Stiftung} (grant P-LS-SPII/18) and from the German {\em Bundesministerium f\"ur Bildung and Forschung} via the ASTRONET project STAR FORMAT (grant 05A09VHA). R.S.K. furthermore gives thanks for subsidies from the Deutsche Forschungsgemeinschaft (DFG) under grants no.\ KL 1358/10 and KL 1358/11 as well as via the Sonderforschungsbereich SFB 881 {\em The Milky way System}.

\label{lastpage}

\end{document}